\documentclass[aps,prd,onecolumn,showpacs,eqsecnum,nofootinbib]{revtex4}

\usepackage{epsfig}
\usepackage{graphicx}
\usepackage{dcolumn}
\usepackage{amsmath}
\usepackage{enumerate}

\def\beq{\begin{equation}}
\def\eeq{\end{equation}}

\def\gtap{\mathrel{ \rlap{\raise 0.511ex \hbox{$>$}}{\lower 0.511ex
   \hbox{$\sim$}}}} 
\def\ltap{\mathrel{ \rlap{\raise 0.511ex
    \hbox{$<$}}{\lower 0.511ex \hbox{$\sim$}}}}
\newcommand{\deltaunodue}{\mbox{$\Delta m_{21}^2 $}}
\newcommand{\deltaunotre}{\mbox{$\Delta m_{31}^2 $}}

\newcommand{\probdiff}{\mbox{${\cal{D}}$}}
\newcommand{\nova}{\mbox{NO$\nu$A} }
\newcommand{\novasp}{\mbox{NO$\nu$A$\!$} }

\newcommand{\pbarP}{P \hspace{-3mm}
\raisebox{1.9ex}{{\tiny(}}\raisebox{1.4ex}{--}\raisebox{1.9ex}{\tiny)}
   }

\begin{document}

\vskip-6pt \hfill {CERN-PH-TH/2005-195} \\
\vskip-6pt \hfill {IPPP/05/63} \\
\vskip-6pt \hfill {DCPT/05/126} \\
\vskip-6pt \hfill {FERMILAB-PUB-05-461-T} \\

\title{
\vskip-12pt~\\
Determining the Neutrino Mass Hierarchy and CP--Violation 

in \nova with a Second Off-Axis Detector}

\author{
\mbox{Olga Mena$^{1}$},
\mbox{Sergio Palomares-Ruiz$^{2}$} and
\mbox{Silvia Pascoli$^{3,4}$}}

\affiliation{
\mbox{$^1$ Theoretical Physics Department, Fermi National Accelerator
Laboratory, Batavia, IL 60510-0500, USA}
\mbox{$^2$ Department of Physics and Astronomy, Vanderbilt University,
  Nashville, TN 37235, USA}
\mbox{$^3$ Physics Department, Theory Division, CERN, CH-1211 Geneva
23, Switzerland}
\mbox{$^4$ IPPP, Department of Physics, University of Durham, Durham
  DH1 3LE, United Kingdom}
\\
{\tt omena@fnal.gov},
{\tt sergio.palomares-ruiz@vanderbilt.edu},
{\tt Silvia.Pascoli@cern.ch}
}

\begin{abstract}

We consider a Super-\novasp-like experimental configuration based on
the use of two detectors in a long-baseline experiment as \novasp. We
take the far detector as in the present \nova proposal and add a second
detector at a shorter baseline. The location of the second off-axis
detector is chosen such that the ratio $L/E$ is the same for both
detectors, being $L$ the baseline and $E$ the neutrino energy. We
consider liquid argon and water-\v{C}erenkov techniques for 
the second off-axis detector and study, for different experimental
setups, the detector mass required for the determination of the
neutrino mass hierarchy, for different values of $\theta_{13}$. We
also study the capabilities of such an experimental setup for
determining CP--violation in the neutrino sector. Our results show that
by adding a second off-axis detector a remarkable enhancement on the
capabilities of the current \nova experiment could be achieved.

\end{abstract}


\maketitle

\section{Introduction}

During the last several years the physics of neutrinos has achieved a
remarkable progress. The experiments with
solar~\cite{sol,SKsolar,SNO1,SNO2,SNO3,SNOsalt}, 
atmospheric~\cite{SKatm}, reactor~\cite{KamLAND} and recently also
long-baseline accelerator~\cite{K2K} neutrinos have provided
compelling evidence for the existence of neutrino oscillations.
The most significant recent contributions are given by
the new Super-Kamiokande data on the $L/E$ dependence of multi-GeV
$\mu$-like atmospheric neutrino events~\cite{SKdip04}, $L$ being the
distance traveled by neutrinos and $E$ the neutrino energy, and by the
new more precise spectrum data of the KamLAND~\cite{KL766} and K2K
experiments~\cite{K2K}. For the first time these data show the
oscillatory dependence on $L/E$ of the probabilities of neutrino
oscillations in vacuum~\cite{BP69}.

The existence of neutrino oscillations plays a crucial role in our
understanding of neutrino physics as it implies non-zero neutrino
masses and neutrino mixing. The present data requires\footnote{We
  restrict ourselves to a three-family neutrino analysis. The
  unconfirmed LSND signal~\cite{LSND} cannot be explained within this
  context and might require additional light sterile neutrinos or more
  exotic explanations (see, e.g. Ref.~\cite{LSNDexpl}). The ongoing
  MiniBooNE experiment~\cite{miniboone} is going to test the
  oscillation explanation of the LSND result.} two large
($\theta_{12}$ and $\theta_{23}$) and one small ($\theta_{13}$) angles
in the Pontecorvo--Maki--Nakagawa--Sakata (PMNS) neutrino
mixing matrix~\cite{BPont57}, and at least two mass square differences,
$\Delta m_{ji}^{2} \equiv m_j^2 -m_i^2$, with $m_{j,i}$ the neutrino
masses, one driving the atmospheric neutrino oscillations
($\deltaunotre$) and one the solar ones ($\deltaunodue$). The mixing
angles $\theta_{12}$ and $\theta_{23}$ control the solar and the
dominant atmospheric neutrino oscillations, while $\theta_{13}$ is the
angle limited by the data from the CHOOZ and Palo Verde
experiments~\cite{CHOOZ,PaloV}.

The Super-Kamiokande~\cite{SKatm} and K2K~\cite{K2K} data are well
described in terms of dominant $\nu_{\mu} \rightarrow \nu_{\tau}$
($\bar{\nu}_{\mu} \rightarrow \bar{\nu}_{\tau}$) vacuum
oscillations. The former require the following best fit values of the
mass squared difference and mixing angle: $|\deltaunotre| = 2.1\times
10^{-3}~{\rm eV^2},~\sin^22\theta_{23} = 1.0~$~\cite{SKatm}, whereas
the latter are best explained for $|\deltaunotre| = 2.8\times
10^{-3}~{\rm eV^2},~\sin^22\theta_{23} = 1.0~$~\cite{K2K}. The 90\%
C.L. allowed ranges of these parameters obtained by the
Super-Kamiokande experiment read~\cite{SKatm}:
\beq 
\label{eq:range}
|\deltaunotre| =(1.5 - 3.4)\times10^{-3}{\rm eV^2},~~~~
\sin^22\theta_{23}\geq 0.92.
\eeq

A new preliminary analysis~\cite{suzukitaup} of Super-Kamiokande data,
using a finer momentum binning of multi-GeV neutrinos, yields a
slightly higher best fit value for the mass squared difference,
$|\deltaunotre| = 2.5\times 10^{-3}~{\rm eV^2}$. The allowed
90\%~C.L. interval reads $|\deltaunotre| =(2.0$--$3.0) \times
10^{-3}{\rm eV^2}$. The sign of $\deltaunotre$, $sgn(\deltaunotre)$, 
cannot be determined with the existing data. The two possibilities,
$\deltaunotre > 0$ or $\deltaunotre < 0$, correspond to two different
types of neutrino mass ordering: normal hierarchy (NH), $m_1 <
m_2 < m_3$ ($\deltaunotre > 0$), and inverted hierarchy (IH),
$m_3 < m_1 < m_2$ ($\deltaunotre < 0$). In addition, information on
the octant where $\theta_{23}$ lies, if $\sin^22\theta_{23} \neq 1.0$,
is beyond the reach of present experiments.

The 2-neutrino oscillation analysis of the solar neutrino data,
including the results from the complete salt phase of the Sudbury
Neutrino Observatory (SNO) experiment~\cite{SNOsalt}, in combination
with the recent KamLAND 766.3 ton-year spectrum data~\cite{KL766},
shows that the solar neutrino oscillation parameters lie in the low-LMA
(Large Mixing Angle) region, with best fit values~\cite{SNOsalt}
\beq
\deltaunodue =8.0 \times 10^{-5}~{\rm eV^2},~~~
\sin^2 \theta_{12} =0.31.
\eeq

A combined 3-neutrino oscillation analysis of the solar, atmospheric,
reactor and long-baseline neutrino data gives~\cite{GG04} (see also
Ref.~\cite{MSTV04}) constrains the third mixing angle to be:
\beq
\sin^2\theta_{13} < 0.041,~~~~3\sigma~{\rm C.L.}
\label{eq:chooz}
\eeq

The future goals in the study of neutrino properties will be to
precisely determine the already measured oscillation parameters
and to obtain information on the unknown ones, namely $\theta_{13}$,
the CP--violating phase $\delta$ and the type of neutrino mass
hierarchy (or equivalently $sgn(\deltaunotre)$). A more accurate
measurement of the leading neutrino oscillation parameters will be
achieved by the MINOS~\cite{MINOS}, OPERA~\cite{OPERA} and
ICARUS~\cite{ICARUS} experiments and future atmospheric and solar
neutrino detectors~\cite{SAWG}. The determination of the $\theta_{13}$
angle is crucial as it opens up the possibility of the experimental
measurement of the CP-- (or T--) violating phase $\delta$, and to
establishing the type of neutrino mass hierarchy. A wide experimental
program is under discussion in order to achieve these goals. 

The mixing angle $\theta_{13}$ controls the $\nu_\mu \rightarrow
\nu_e$ and $\bar{\nu}_\mu \rightarrow \bar{\nu}_e$ conversions in
long-baseline experiments and the $\nu_e$ disappearance in
short-baseline reactor experiments. Present and future conventional
beams~\cite{MINOS}, super-beams with an upgraded proton source and a
more intense neutrino flux~\cite{newNOvA,T2K}, and future reactor 
neutrino experiments~\cite{futurereactors} have the possibility to
measure, or set a stronger limit on, $\theta_{13}$. Smaller values of
this mixing angle could be accessed by very long baseline experiments
such as neutrino factories~\cite{nufact,CDGGCHMR00}, or by
$\beta$-beams~\cite{zucchelli,mauro,BCCGH05,betabeam} which exploit
neutrinos from boosted-ion decays, or by super-beams with Megaton
detectors~\cite{T2K}, or by electron-capture
facilities~\cite{ecapt}. The mixing angle $\theta_{13}$ also controls
the Earth matter effects in multi-GeV
atmospheric~\cite{atmmatter1,mantle,core,atmmatter2,atmmatter3} and in 
supernova~\cite{SN} neutrino oscillations.

The magnitude of the T-violating and CP--violating terms as well as
matter effects in long baseline neutrino oscillations is controlled by
$\sin\theta_{13}$~\cite{CPT}. If the value of $\theta_{13}$ is sizable
and at the reach of future experiments, it would be possible to search
for CP--violation in the lepton sector and to determine the type of
neutrino mass hierarchy by establishing
$sgn(\deltaunotre)$~\footnote{Information on the type of neutrino mass
  hierarchy might be obtained in future atmospheric neutrino
  experiments~\cite{mantle,core,atmmatter2,atmmatter3}. If neutrino
  are Majorana particles, next generation of neutrinoless double
  $\beta-$decay experiments could establish the type of neutrino mass
  spectrum~\cite{PPRmass} (see also
  Refs.~\cite{Bilmass,BPP,PPW,betabetamass}) and, possibly, might
  provide some information on the presence of CP--violation in the
  lepton sector due to Majorana CP--violating phases~\cite{PPRCP} (see
  also Refs.~\cite{BilCPV,BPP,PPW,betabetaothers}).}. Typically, the
proposed experiments have a single detector and plan to run with the
beam in two different modes, neutrinos and antineutrinos. In
principle, by comparing the probability of neutrino and antineutrino
flavor conversion, the values of the CP--violating phase $\delta$ and
of $sgn(\deltaunotre)$ can be extracted. Different sets of values
of CP--conserving and violating parameters, $(\theta_{13}, \theta_{23},
\delta, sgn(\deltaunotre))$, lead to the same probabilities of
neutrino and antineutrino conversion and provide a good description of
the data at the same confidence level. This problem is known as the
problem of degeneracies in the neutrino parameter
space~\cite{FL96,BCGGCHM01,MN01,BMWdeg,deg} and severely affects the
sensitivities to these parameters in future long-baseline experiments.
Many strategies have been advocated to resolve this issue. Some of the
degeneracies might be eliminated with sufficient energy or baseline
spectral information~\cite{CDGGCHMR00}. However, statistical errors
and realistic efficiencies and backgrounds limit considerably the
capabilities of this method. Another
detector~\cite{BNL,MN97,BCGGCHM01,silver,BMW02off,MPP05,twodetect} or
the combination with another
experiment~\cite{BMW02,HLW02,MNP03,otherexp,mp2,HMS05,M05} would,
thus, be necessary~\footnote{New approaches which exploit other
  neutrino oscillations channels such as muon neutrino disappearance
  have been proposed~\cite{hieratm} for determining the type of
  hierarchy. They require very precise neutrino oscillation
  measurements. Under certain rather special conditions it might be
  determined also in experiments with reactor
  $\bar{\nu}_e$~\cite{SPMPiai01}.}.

The use of only a neutrino beam could help in resolving the type of
hierarchy when two different long-baselines are
considered~\cite{HLW02,MNP03,MPP05,twodetect}. The most favorable case
is keeping the same $L/E$ at the two different baselines. In
particular, it was shown that just one experiment, named
Super-\novasp~\cite{MPP05}, which runs in the neutrino mode and uses
two detectors at different distances and different off-axis angles,
could determine the type of hierarchy for values of $\sin^2
\theta_{13}$ as small as 0.02. This method is free of degeneracies. It
uses the fact that, comparing the probability of conversion at the two
sites, the vacuum oscillation term cancels out, leaving as dominant
the term sensitive to matter effects. Hence the leading CP--violating
term cancels and CP--violation gives only a subdominant
correction. Off-axis neutrino beams have a very narrow neutrino 
spectra, and their peak energy can be tuned by displacing the detector 
out of the main beam axis. We notice that an off-axis beam can be
obtained by either putting the detector a few km away from the
location of an on-axis surface detector, or by placing it on the
vertical of the beam-line but at a much shorter distance. In such a
way, a single beam could do the job of two beams with different
energies. In Ref.~\cite{MPP05} the case of the NuMI beam and two
50~kton liquid argon TPC detectors, one at the \nova proposed
site~\cite{newNOvA} and the second at a shorter baseline, 200 and
434~km, was studied in detail. It was shown that the hierarchy can be 
determined at 95~\%~C.L., regardless of the value of $\delta$, for
$\sin^2 \theta_{13} \geq 0.05$ for a conventional beam and for $\sin^2
\theta_{13} \geq 0.02$ with a proton driver. This should be compared
with the sensitivity of the proposed \nova
experiment~\cite{newNOvA}. At 95~\%~C.L., only for less than 40~\% of
the values of $\delta$, the type of neutrino mass hierarchy can be
resolved if $\sin^2 2 \theta_{13} \leq 0.10$ in a 3 neutrino plus 3
antineutrino running~\cite{newNOvA}.

Here, we follow the Super-\nova strategy~\cite{MPP05} but we
consider a more realistic scenario. We consider the use of the
proposed \nova detector, a 30~kton low-Z calorimeter, at the far
site. We add a second off-axis detector, either a liquid argon or a
water-\v{C}erenkov detector, at a shorter baseline of 200~km, keeping
$L/E$ constant at the two sites. In addition, we provide a sequencing
for the construction of the experiment, assuming that the second
detector will be constructed at a second stage. If the mixing angle
$\theta_{13}$ is very small, then adding the second detector would
increase the statistics for constraining this parameter but would not
provide any useful information on the type of neutrino mass
hierarchy. On the contrary, if $\theta_{13}$ is within the sensitivity
of the \nova experiment and a positive signal is found in the first
years of running, the construction of the second off-axis detector
would enormously enhance the capabilities for the determination of the
type of neutrino mass hierarchy, (almost) free of degeneracies. Let us
stress that the unique contribution of the NuMI neutrino program and
of the \nova experiment would be establishing the type of
hierarchy~\footnote{The proposed atmospheric neutrino experiment
  INO~\cite{ino} could achieve a similar result on the same timescale
  if finally approved.}. Therefore we focus on this issue and we study
in detail the requirements for the second detector for achieving this
remarkable result. In addition, we also study the possibility of the
measurement of CP--violation within this experimental setup. In order
for this to be possible, the running with antineutrinos is completely
necessary. Hence, in our sequencing of the experimental setup, we also
add some years of antineutrino data.  

We start by reviewing the relevant formalism and the power of the
method in Sec.~\ref{formalism}. In Sec.~\ref{setup} we describe the two
different experimental configurations we consider. We show in
Sec.~\ref{matter} the results for different possibilities for the
200~km detector and we depict how the sensitivity changes for
different values of $|\deltaunotre|$. In Sec.~\ref{diffloc}, we
compare different possible choices for the location of the second
off-axis detector and argue why the choice we make here is
preferable. Finally, in Section.~\ref{conclusions}, we summarize and
conclude.

\section{Formalism}
\label{formalism}

In the present study we focus on the capabilities of reducing
degeneracies of a long-baseline neutrino experiment with two off-axis 
detectors. It is well known that an off-axis neutrino beam is
characterized by a well peaked spectrum, so for sake of
analytical understanding we can take it as approximately
monochromatic. It was shown in Refs.~\cite{MPP05,MNP03,HLW02} that a
particularly suitable configuration for the determination of the
neutrino mass hierarchy is such that the ratio $L/E$, where $L$ is the
baseline and $E$ is the neutrino energy, is the same for both
detectors. For this particular configuration, the use of just a
neutrino running could be sufficient to determine the type of neutrino
mass spectrum, depending upon the actual value of $\sin^2
2\theta_{13}$. We review below the analytical expressions illustrating
this result.

For neutrino energies $E \gtap$ 1 GeV, $\theta_{13}$ within
the present bounds~\cite{GG04,MSTV04}, and baselines $L \ltap
1000$~km~\cite{BMWdeg}~\footnote{For $E \gtap 0.6$ GeV we have checked
  that the analytical expansion is accurate for $L < 500$ km within the
  present bounds of $\theta_{13}$ and
  $\deltaunodue/\deltaunotre$~\cite{MPP05}.}, the oscillation
probability $\pbarP(L)$ can be expanded in the small parameters
$\theta_{13}$, $\Delta_{12}/\Delta_{13}$, $\Delta_{12}/A$ and
$\Delta_{12} L$ , where $\Delta_{12} \equiv \deltaunodue/(2 E)$ and
$\Delta_{13} \equiv \deltaunotre/(2E)$~\cite{CDGGCHMR00} (see also
Ref.~\cite{3prob}):
\beq
\begin{array}{ll}
\pbarP(L) \simeq &
\sin^2 \theta_{23} \, \sin^2 {2 \theta_{13} } \left(
\frac{\Delta_{13}}{A \mp \Delta_{13}} \right)^2
\sin^2 \left( \frac{(A \mp \Delta_{13}) L}{2} \right) \\
& + \cos \theta_{13} \sin {2 \theta_{13} } \sin {2 \theta_{23}} 
\sin {2 \theta_{12}} \ \frac{\Delta_{12}}{A} \frac{\Delta_{13}}{A \mp
  \Delta_{13}} \
\sin \left(\frac{A L}{2} \right) \sin \left( \frac{(A \mp
  \Delta_{13}) L}{2} \right)
 \cos \left(\frac{\Delta_{13} L}{2} \mp \delta \right) \\
 & + \cos^2 \theta_{23} \sin^2 {2 \theta_{12}} \left(
 \frac{\Delta_{12}}{A} \right)^2 \sin^2 \left( \frac{A L}{2} \right).
\end{array}
\label{eq:probappr}
\eeq

Here, we use the constant density approximation for the index of
refraction in matter $A \equiv \sqrt{2} G_{F} \bar{n}_e(L)$. We define
$\bar{n}_e(L)= 1/L \int_{0}^{L} n_e(L') dL'$ as the average electron
number density, with $n_e(L)$ the electron number density along the
baseline.  In addition, taking into account that for the baselines and
energies of interest the matter effects are small ($A \ll
\deltaunotre$ and $AL \ll 1$) a further expansion can be carried
out. It was shown in Ref.~\cite{MPP05} that for this experimental
configuration of the detectors, at first order, the $\nu_\mu
\rightarrow \nu_e$ conversion probability in vacuum is the same at
both locations, so the dominant term of the normalized difference of
the oscillation probabilities computed at the near and far baselines,
$\probdiff \equiv \frac{P(L_{\rm N})-P(L_{\rm F})}{P(L_{\rm
    N})+P(L_{\rm F})}$, depends linearly on the matter term. Up to
terms of order $\mathcal{O}(A \, \Delta_{12}/\Delta_{13})$,
$\mathcal{O}(A \, \theta_{13}^2)$ and $\mathcal{O}(A^2)$, the
expression for $\probdiff$ reads~\cite{MPP05}:
\begin{eqnarray}
\probdiff  \! &  \! \simeq  \! & \!
\frac{A_{\rm N} L_{\rm N} \!  - \!  A_{\rm F} L_{\rm F}}{2} \!
   \left( \frac{1}{(\Delta_{13} L / 2)} - \frac{1}{\tan(\Delta_{13} L
  / 2)} \right) \!
 \left( \! 1- \frac{\Delta_{12}}{\Delta_{13}} \frac{\cos \theta_{13}}
   {\tan \theta_{23}} \ \frac{\sin {2 \theta_{12}}}{\sin 2
   \theta_{13}} \ \frac{\Delta_{13} L / 2}{ \sin (\Delta_{13} L / 2)}
   \cos \left(\!  \delta \! + \! \frac{\Delta_{13} L}{2} \!  \right)
 - \frac{\sin^2 2 \theta_{13} }{2} \! \nonumber \right) \\[2ex]
 & + &
  \frac{1}{2} \frac{A_{\rm N}^2 L_{\rm N}^2 - A_{\rm F}^2 L_{\rm
      F}^2}{4} \left( 
  \frac{1}{(\Delta_{13} L / 2)^2} - \frac{1}{\sin^2 \left( \Delta_{13}
   L / 2 \right)} \right)~.
\label{eq:probdifffull}
\end{eqnarray}

As shown in this formula, the corrections due to the CP--violating
terms are small as far as $\sin^2 \theta_{13} \gtap 0.02$ because
the dominant CP--violating terms in $P(L_{\rm N})$ and $P(L_{\rm F})$,
which depend upon the vacuum terms, cancel out. As we can clearly see
from Eq.~(\ref{eq:probdifffull}), by considering both detectors with
the same vacuum oscillation phase, the dependence on the matter
potential dominates, with
little contamination from other parameters. Thus, the normalized
asymmetry is sensitive to the type of hierarchy as the sign of the
matter potential term is a direct indication of the sign of the
atmospheric mass squared difference.  This is the power of the method,
for corrections due to the rest of parameters can modify slightly its
value, but cannot change its sign.

In Ref.~\cite{MPP05}, the determination of $sgn(\deltaunotre)$ was
considered and studied in detail with a configuration based on the
(old) \nova proposal~\cite{NOvA} with a single running in the neutrino
mode. Although, as was shown there, this could be the right way to
proceed for the resolution of the mass hierarchy, running in
antineutrinos is necessary in order to search for effects due to the
CP--violating phase $\delta$. Hence, in the present study we consider
the possibility to run also in the antineutrino channel and we analyze
to what extent this opens the possibility to search also for
CP--violation in the lepton sector. In an analogous way as for the
previous case, we can compute the quantity $\probdiff_{CP} \equiv
\frac{P(L)-\bar{P}(L)}{P(L)+\bar{P}(L)}$, i.e., the CP asymmetry for
neutrinos and antineutrino detected at a distance $L$. Up to second
order, $\probdiff_{CP}$ is given by: 
\begin{eqnarray}
\nonumber
\probdiff_{CP} & \simeq & A L \ 
  \left( \frac{1}{\Delta_{13} L / 2} - \frac{1}{\tan(\Delta_{13} L /
2)} \right)
+ \frac{\Delta_{12}}{\Delta_{13}} \frac{\cos \theta_{13}}
{\tan \theta_{23}} \frac{\sin 2 \theta_{12}}{\sin 2 \theta_{13}}
\frac{\Delta_{13} L}{2} \, \biggl[ \, 2 \sin \delta - \\
& - &
\left( \frac{1}{\Delta_{13} L / 2} - \frac{1}{\tan(\Delta_{13} L /
2)} \right)
\frac{1}{\tan(\Delta_{13} L /2)} \ A L
\cos \delta \, \biggr].
\label{eq:probCPsame}
\end{eqnarray}

The main feature of Eq.~(\ref{eq:probCPsame}) lies on the fact that,
for the energies and baselines of interest, the first two terms in the
right-hand-side, the dominant matter effect term and the CP--violating
one, are of the same order, with the latter becoming more important as
$\sin^2 \theta_{13}$ decreases. This typically introduces an important
degeneracy and limits the sensitivity to CP--violation in
long-baseline experiments and is the main reason why having two beams
of neutrino with different $L$ and $E$, but the same $L/E$, gives
better sensitivity to $sgn(\deltaunotre)$ than one neutrino and one
antineutrino beam at the same $L$ and $E$. Hence, by placing the
second detector in such a configuration, as it will be shown below,
the fact that the type of hierarchy could be determined by using the
neutrino channel only, solves this degeneracy and improves the
sensitivity to CP--violation when running with antineutrinos.

\section{Experimental configuration}
\label{setup}

As has been mentioned above, the idea of exploiting the capabilities
of two off-axis detectors, with the same $L/E$ at a single experiment,
was presented in Ref.~\cite{MPP05} for the first time by using the
proposed configuration of the \nova experiment. In that study it was
assumed the use of two liquid argon detectors of 50~kton each, one
placed off-axis at 810~km (the proposed site for \novasp) and the
other one off-axis at 200~km (or 434~km), which was named
Super-\novasp. It was shown that the best capabilities for determining
the type of neutrino mass hierarchy were achieved for the case of a
near off-axis detector placed at the shortest baseline, i.e., 200~km,
for which the differences in the matter effects with respect to the
far detector were larger, and consequently so was the sensitivity to
this parameter. Although the development of the liquid argon
technology is already quite advanced and has been successfully proven
to work in small prototypes~\cite{ICARUS}, it is reasonable to think
that the timelines for the \nova experiment might not allow to build
such a detector at the far site. Hence, in the present study, keeping
the essence of Super-NO$\nu$A, we consider the use of the currently
proposed detector~\cite{newNOvA} at the far site, that is, a low
density tracking calorimeter of 30~kton with an efficiency of 24\%,
and the possibility of building a second off-axis detector at 200~km. 

For the detector technologies we consider water-\v{C}erenkov and
liquid argon and study the case of different sizes. For the
purposes of this study, we consider flat efficiencies with no
energy dependence, $\epsilon$, and backgrounds coming only from the
beam. Consequently, one of the relevant statistical parameters is what 
we define as the \textit{efficient mass of the detector}, $M_{\rm efc}
= M_{\rm{det}} \times \epsilon$, where $M_{\rm{det}}$ is the fiducial
mass of the detector. The proposed \nova detector has $M_{\rm efc}
= 7.2$~kton. We will consider the medium energy configuration of the
NuMI beam and $6.5 \times 10^{20}$ protons on target per year
(pot/yr)~\cite{newNOvA}. In addition, we propose to sequence the
experiment such that the first phase uses only the far detector 
as in the \nova proposal~\cite{newNOvA}. In the second phase, the
second detector, the near off-axis one, is already constructed and
starts taking data. The choice of adding the second detector should be
guided by the knowledge of the value of the mixing angle
$\theta_{13}$. Such information can be provided by the \nova
experiment itself in the first years of running~\cite{MP05} and by
other experiments which are expected to run
previously~\cite{futurereactors} or at the same time~\cite{T2K}. If
the angle $\theta_{13}$ is beyond the reach of the \nova experiment,
adding a second detector would not be useful in determining the type
of neutrino mass hierarchy or CP--violation. Conversely, for a sizable
value of $\theta_{13}$, we show that constructing a near off-axis
detector with the same $L/E$ would allow to determine
$sgn(\deltaunotre)$, (almost) free of degeneracies.

\begin{figure}[t]
\begin{center}
\epsfig{file=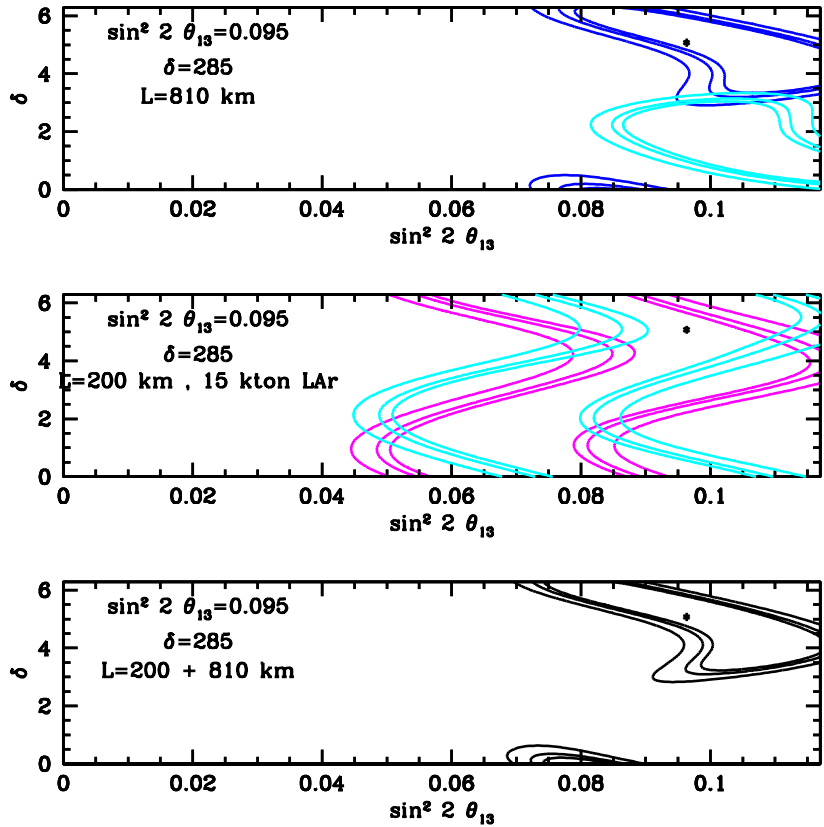,width=12.0cm}
\caption{\textit{$90\%, 95\%$ and $99\%$ C.L. contours resulting from
    the simultaneous extraction of  $\sin^2 2 \theta_{13}$ and
    $\delta$. The true value that we have assumed is $\sin^2 2
    \theta_{13}=0.09$ and $\delta=285^{\circ}$ (denoted by a
    star). The top panel shows the results for the analysis of data
    from the \nova far detector. The blue lines denote the $90\%,
    95\%$ and $99\%$ C.L. contours resulting from the analysis of the
    data assuming normal hierarchy. If the data analysis is performed
    with the opposite sign of the atmospheric mass splitting, fake
    solutions associated to the wrong choice of the neutrino hierarchy
    appear, shown as cyan contours. The medium panel shows the results
    from the analysis of the data at a 15~kton liquid argon detector
    located at 200~km from the neutrino source with magenta
    contours. Again fake solutions are depicted with cyan
    contours. The bottom panel shows the results from the combination
    of the data from the two detectors with black contours. The case
    considered is that of scenario I (\textbf{ScI}) for which the far
    detector is assumed to take data during 9 years of neutrino
    running and 5 years of antineutrino running, while the near
    detector is assumed to take data during 6 years of neutrino
    running and 2 years of antineutrino running. The total number of
    years would be 14. Let us notice that the choice of the true value
    of the $\delta$ phase is the least favorable one. For different
    values of $\delta$ a better sensitivity would be achieved.}}
\label{fig:0.09}
\end{center}
\end{figure}

The \nova experiment itself has the capability to probe values of
$\theta_{13}$ down to $\sin^2 \theta_{13}\sim 0.02$, within the first
3 years of running~\cite{MP05}. Therefore, we consider first only the
\nova far detector and 3 years of neutrino run plus 3 years of
antineutrino run. Assuming that a positive signal for $\nu_\mu 
\rightarrow \nu_e$  conversion is found after the first neutrino run,
signaling $\theta_{13}$ in the interesting range of values, we assume
that the second detector is constructed and starts taking data
immediately after the antineutrino running period.

Following these general guidelines we have considered two different
scenarios. In scenario I (\textbf{ScI}), without proton driver, after
the first 6 years with only the \nova detector, we assume other 6
years of neutrino run plus 2 years of antineutrino run. We do this for
different values of $M_{\rm efc} = 17.5, 35, 70, 13.5, 27, 45$~kton,
where the first three values would account for a water-\v{C}erenkov
detector (70\% efficiency) of 25, 50 and 100~kton, whereas the last
three values would account for a liquid argon detector
(90\% efficiency) of 15, 30 and 50 kton. In scenario II
(\textbf{ScII}), the beam is upgraded from the beginning with a proton 
driver ($25 \times 10^{20}$ pot/yr), which is equivalent to multiply
$M_{\rm efc}$ by $\sim$ 3.85. During the first 6 years, only the \nova
detector would be taking data. Afterwards, both near and far detectors
would take data for 3 years of neutrino run plus 1 year of antineutrino
run. We use the same values of $M_{\rm efc}$ as for \textbf{ScI}. The
total number of years in \textbf{ScI} (\textbf{ScII}) is of 14 (10).

\begin{figure}[t]
\begin{center}
\epsfig{file=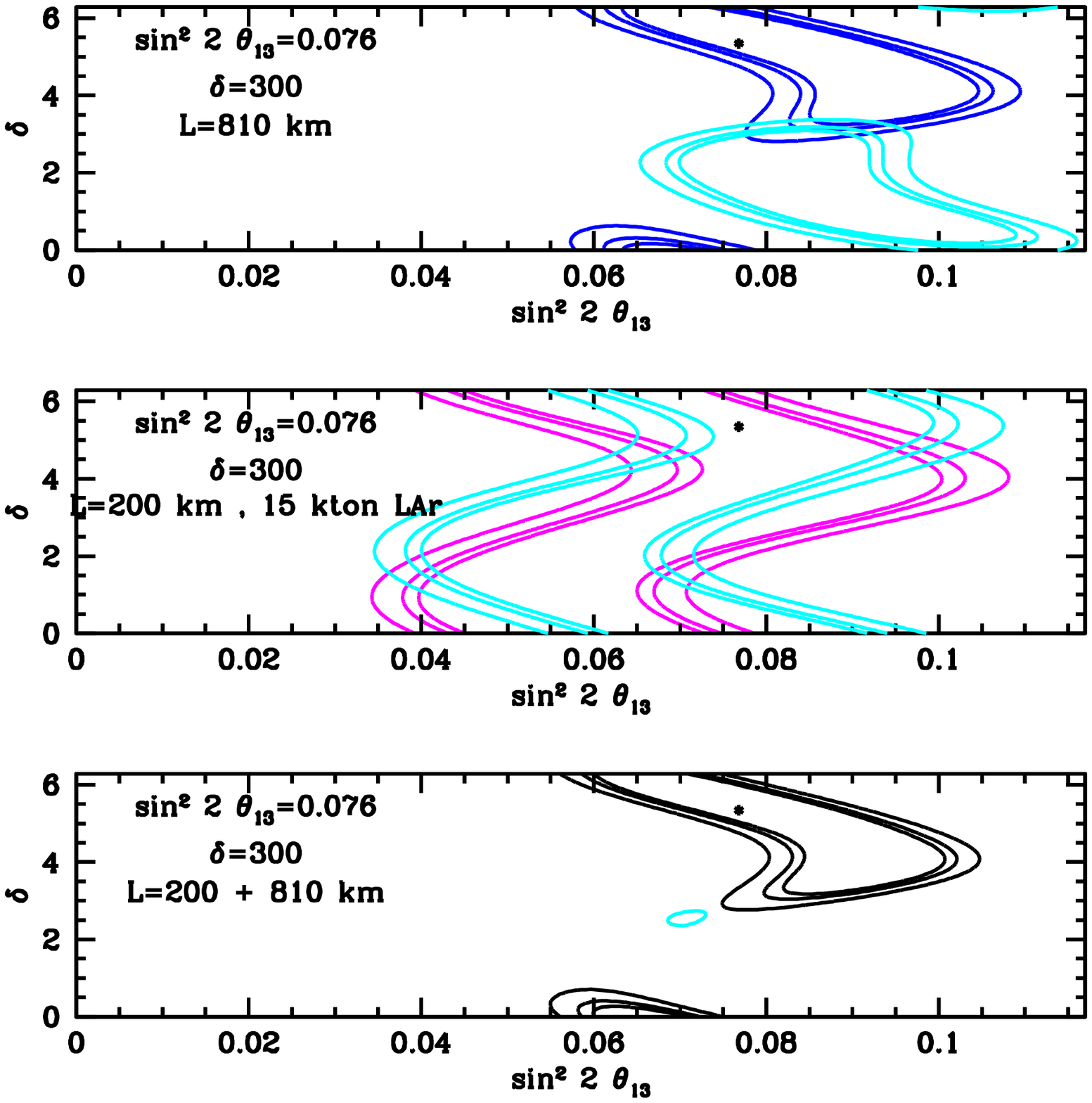,width=12.cm} 
\caption{\textit{ The same as in Fig.~(\ref{fig:0.09}) but for $\sin^2
    2 \theta_{13}=0.076$ and $\delta=300^{\circ}$.}}
\label{fig:0.07}
\end{center}
\end{figure}

The question that we would like to address is what is the minimum
\textit{efficient mass} of the near off-axis detector required to
ensure the resolution of the neutrino mass hierarchy for the full
range of values of the CP--violating parameter $\delta$ for a given
value of  $\theta_{13}$. We have thus performed a $\chi^2$ analysis
for both $\theta_{13}$ and $\delta$. For a given value of the
oscillation parameters, we have computed the expected number of
electron events, $N_{\ell}$, detected at the possible locations $\ell
= {\rm N, F}$ (near,far sites). The observable that we exploit,
$N_{\ell}$, is given by 
\beq
N_{\ell,\pm} = \int^{E_{\rm max}}_{E_{\rm min}} \;
\Phi_{\ell,\nu}(E_\nu,L) \; \sigma_{\nu}(E_\nu) \;
 P_{\nu}(E_\nu, L, \theta_{13}, \delta, \deltaunotre,
{\alpha}) \; dE_{\nu}
\eeq
where the sign +($-$) applies for the normal (inverted) hierarchies and
${\alpha}$ is the set of remaining oscillation parameters:
$\theta_{23}, \, \theta_{12}, \, \deltaunodue$ and the matter parameter
$A$ (which depends on the baseline under consideration), which are
taken to be known; $\Phi_{\ell,\nu}$ denotes the neutrino flux,
$\sigma_{\nu}$ the relevant cross sections and $P_\nu$ the $\nu_\mu
\rightarrow \nu_e$ conversion probability. The convolution of these
three magnitudes are then integrated over a narrow energy window of
$1$ GeV, where $E_{\rm min}$ and $E_{\rm max}$ refer to the lower and
upper energy limits, respectively. We have divided the total number
of events in two bins of equal width $\Delta E_\nu=0.5$
GeV~\footnote{This is a very conservative estimate for the neutrino
  energy resolution, which in the range of interest is $\Delta E_\nu
  /E_\nu \sim 50\%$.}. 

For our analysis, unless otherwise stated, we will use a
representative value of $|\deltaunotre| = 2.4 \times 10^{-3} \
\rm{eV}^2$, which lies within the best-fit values for the
Super-Kamiokande~\cite{SKatm} and K2K~\cite{K2K} experiments. However,
we will also present how the $\chi^2$ analysis results depend on the
value of the atmospheric mass difference $|\deltaunotre|$.
For the remaining oscillation parameters, $\theta_{23}$,
$\deltaunodue$ and $\theta_{12}$, we will use the best fit values
quoted in the introduction.

\section{Determination of the type of hierarchy and of CP--violation}
\label{matter}

For the progress in the studies of neutrino physics it is of
fundamental importance measuring the value of the small $\theta_{13}$
angle, determining the type of neutrino mass hierarchy and searching
for possible CP--violation in the neutrino sector. Possibly, the only
experiments available to be sensitive to all these three parameters in
the near future are long-baseline experiments. However, the chances
for them to succeed depend upon the value of $\theta_{13}$. If the 
mixing angle $\theta_{13}$ is out of their sensitivity reach, then
next generation of neutrino experiments would be needed, such as
neutrino factories~\cite{nufact,CDGGCHMR00}, $\beta$-beams or
super-beams~\cite{zucchelli,mauro,BCCGH05,betabeam} or
electron-capture facilities~\cite{ecapt}. In this study we assume that
the value of $\theta_{13}$ is within the sensitivity of a
long-baseline experiment like \nova and of future reactor neutrino
experiments~\cite{futurereactors}, that is $\sin^2 2\theta_{13} \gtap
0.01$. With this assumption, we tackle the problem of resolving the
type of neutrino mass hierarchy and determining CP--violation by
considering the experimental setup described above. We show that the
improvement with respect to the current \nova proposal is truly
remarkable.

\begin{figure}[t]
\begin{center}
\epsfig{file=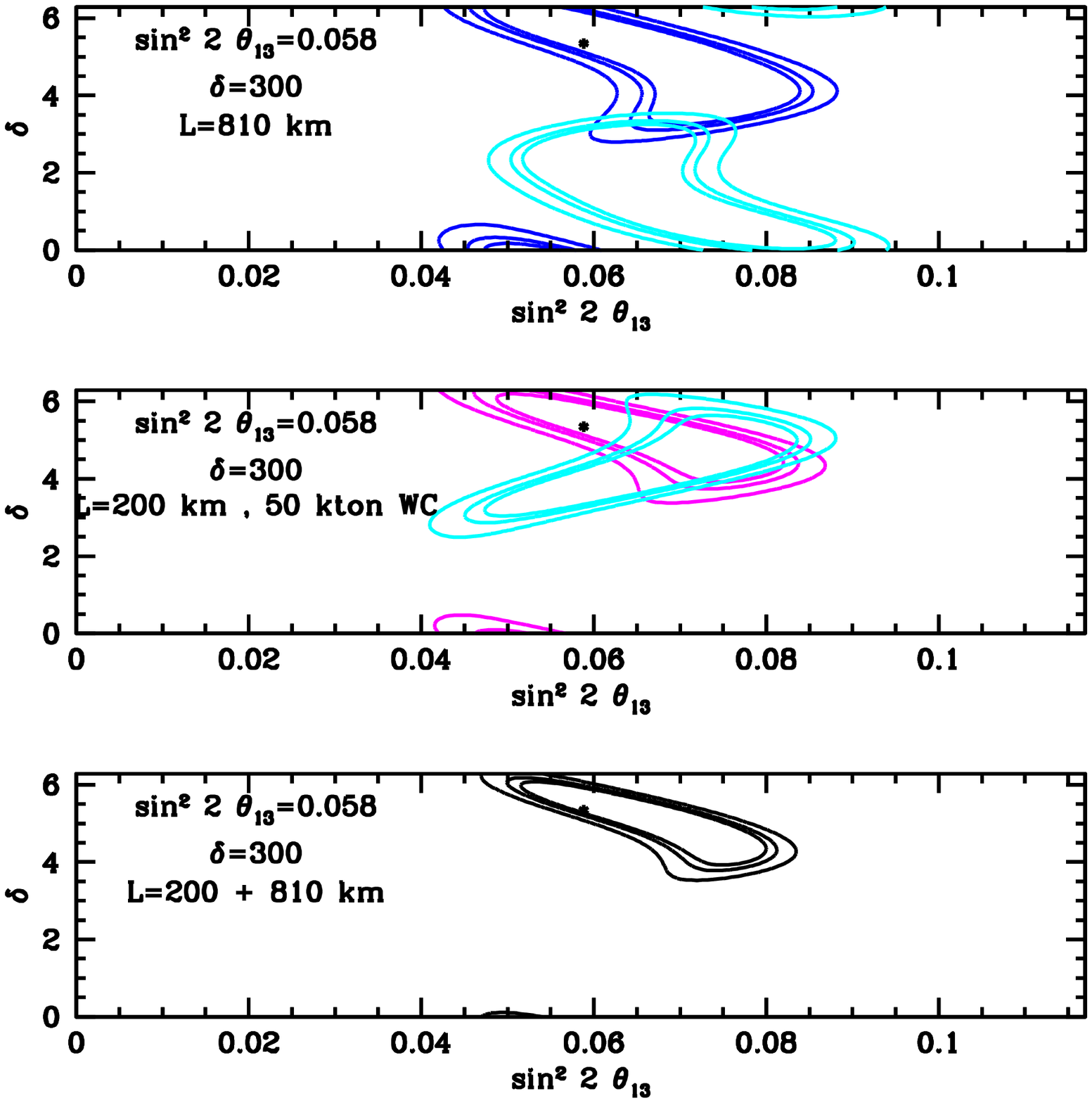,width=12.cm}
\caption{\textit{ The same as in Fig.~(\ref{fig:0.09}) but for $\sin^2
    2 \theta_{13}=0.058$ and $\delta=300^{\circ}$. The near detector
    has been upgraded to a 50 kton water-\v{C}erenkov detector.}}
\label{fig:0.058}
\end{center}
\end{figure}

\begin{figure}[t]
\begin{center}
\epsfig{file=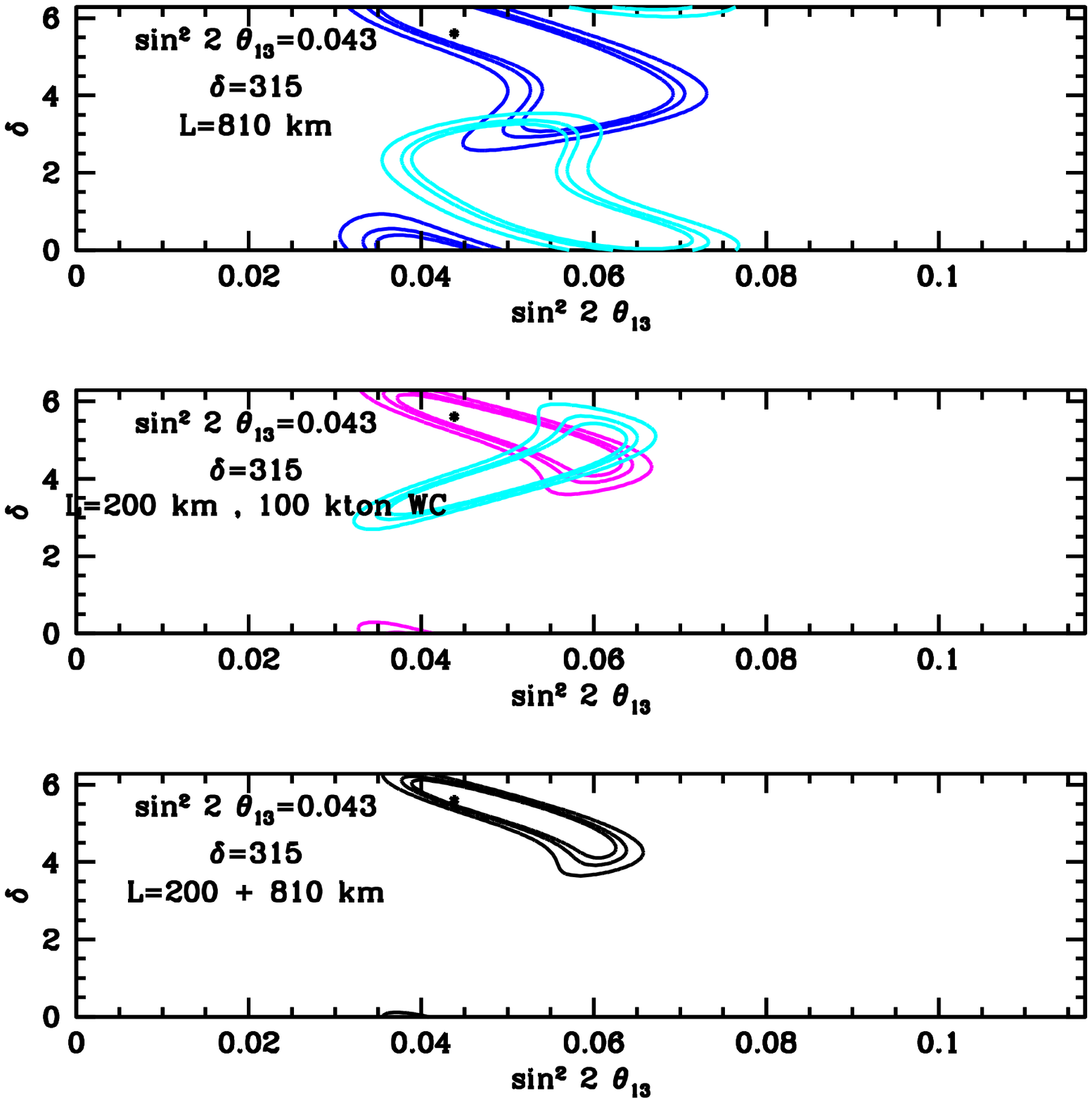,width=12.0cm}
\caption{\textit{ The same as in Fig.~(\ref{fig:0.09}) but for $\sin^2
    2 \theta_{13}=0.043$ and $\delta=315^{\circ}$. The near detector
    has been upgraded to a 100 kton water-\v{C}erenkov detector.}}
\label{fig:0.043}
\end{center}
\end{figure}

We first explore what is the minimal scenario which could provide 
the mass hierarchy resolution for different values of $\sin^{2} 2
\theta_{13}$. For doing this, we have performed a $\chi^{2}$ analysis
of the data in the ($\sin^{2} 2 \theta_{13}$, $\delta$) plane. We
assume Nature has chosen the normal or inverted hierarchy and we
attempt to fit the data to the expected number of events for the
opposite hierarchy. Generically one expects two fake solutions
associated with the wrong choice of the hierarchy at fixed neutrino
energy and baseline. The $\chi^2$ function in the combination of two
baselines and of the neutrino and antineutrino channels reads
\beq
\chi_{\ell \ell'}^2 = \sum_{\ell \ell'} \sum_{p = e^+,e^-} \;
(\mathcal{N}_{\ell,\pm} - N_{\ell,\pm}) C_{\ell:\ell'}^{-1}
(\mathcal{N}_{\ell',\pm} - N_{\ell',\pm})\,, 
\label{chi2c}
\eeq
where the + ($-$) sign  refers to normal (inverted) hierarchy and $C$
is the covariance matrix, which for the particular analysis considered
in the present study, contains only statistical errors.
Systematic errors due to the solar neutrino parameters
can be safely neglected, while those for the atmospheric mixing
parameters $\deltaunotre$ and $\sin^2 2 \theta_{23}$ will be at
the level of $5\%$ and $2\%$ respectively~\cite{newNOvA,messierp04}
when \nova is expected to start taking data. On the other hand, the
input particle production spectra from which the neutrino ones at the
detectors are simulated are only known to about the 20\% level at
present. Before \nova starts taking data, we expect the MIPP
experiment to improve this knowledge to about the 5\%
level~\cite{MIPP}. By using the two detectors, whose flux can be
related by simple kinematics in a Monte Carlo simulation, this
systematic error would be further reduced. Our knowledge of CC cross
sections represents another important source of systematic errors. For
the energy range of interest, the uncertainties amount to about
20-30\% at present. However, during the next few years, experiments
like K2K~\cite{k2kcross}, MiniBooNE~\cite{minicross} and
Miner$\nu$a~\cite{minerva} will reduce all these errors substantially,
to the 5\% level. Systematic errors related to the detectors are
expected to be smaller~\cite{flare}. The impact of the uncertainties
related to the beam and cross sections has been found analitically to
be negligible compared to the uncertainties in the oscillation
parameters. All in all, a detailed analysis with a full simulation of
these errors has not been yet performed~\cite{newNOvA} and would be
required. Nevertheless, we are confident on the fact that the impact
of systematic errors should not change our basic conclusions.

The experimental ``data'', ${\mathcal N_{\ell,\pm}}$, are given by
\beq
\mathcal N_{\ell,\pm} = \langle N_{\ell,\pm} + N_{b\ell }\rangle -
N_{b\ell, \pm}\, ,
\eeq
where we have considered that the efficiencies are flat in the visible
energy window, $N_{b\ell }$ are the background events and $\langle
\rangle$ means a Gaussian/Poisson smearing (according to the
statistics).

\begin{figure}[t]
\begin{center}
\epsfig{file=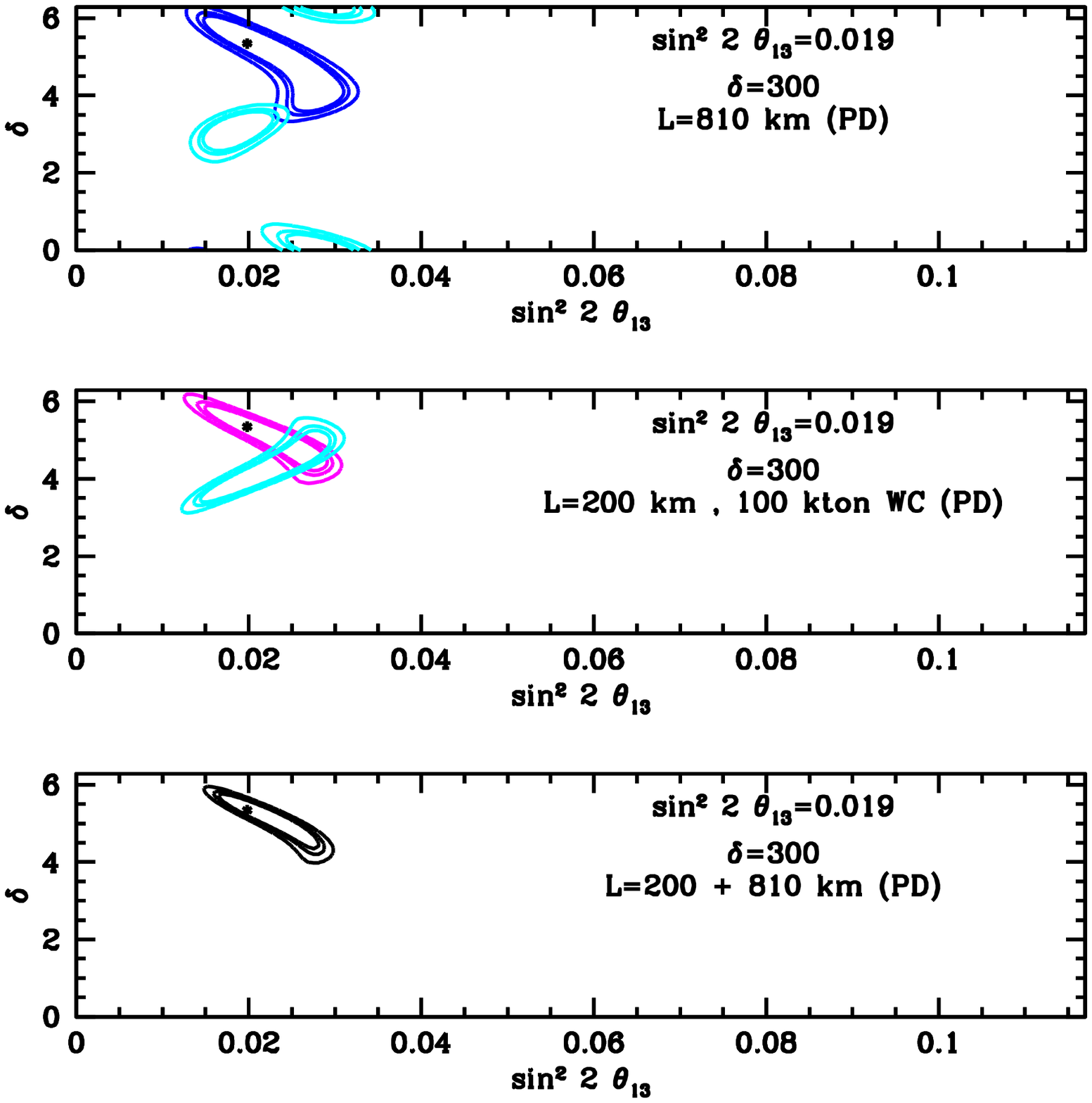,width=12.0cm}
\caption{\textit{The same as in Fig.~(\ref{fig:0.09}) but for $\sin^2
    2 \theta_{13}=0.019$ and $\delta=300^{\circ}$. The near detector
    has been upgraded to a 100 kton water-\v{C}erenkov and the case
    exploited is that of scenario II (\textbf{ScII}) with a proton
    driver. The far detector is assumed to take data during 6 years of
    neutrino running and 4 years of antineutrino running, while the
    near detector is assumed to take data during 3 years of neutrino
    running and 1 year of antineutrino running. The total number of
    years equals 10.}}
\label{fig:0.019pd}
\end{center}
\end{figure}

We start the analysis by assuming that the true value of $\sin^2 2
\theta_{13}$, that is, the value that Nature has chosen, is close
to its present upper bound. We depict in Fig.~\ref{fig:0.09} the
90\%, 95\% and 99\% C.L. contours resulting from the simultaneous
extraction of  $\sin^2 2 \theta_{13}$ and $\delta$. The true values
that we have assumed are $\sin^2 2 \theta_{13}=0.095$ and
$\delta=285^{\circ}$ and the point in the ($\sin^{2} 2 \theta_{13}$,
$\sin \delta$) plane is denoted by a star. The top panel shows the
results for the analysis of data from only the \nova far detector. For
this particular value of $\sin^2 2 \theta_{13}$ we have considered the
scenario I (\textbf{ScI}), with no proton driver. Therefore, the
statistics at the far detector corresponds to 9 years of neutrino plus
5 years of antineutrino running. The blue lines denote the 90\%, 95\%
and 99\% C.L. contours resulting from the analysis to the data
assuming NH. If the data analysis is performed with IH, a fake
solution associated to the wrong choice of the neutrino mass hierarchy
appears. We have depicted these $sgn(\deltaunotre)$-degeneracies by
cyan contours. Notice, from the results depicted if
Fig.~\ref{fig:0.09}, that the future \nova experiment (with just one
far detector) might not be able to determine the neutrino mass
hierarchy if Nature has chosen the central values in
Fig.~\ref{fig:0.09}, even if the statistics is increased by more than
a factor of two.

The medium panel shows the results from the analysis of the data at a
near detector located at 200~km from the neutrino source with magenta
contours. The off-axis location of the near detector is chosen to
ensure the same $L/E$ at the near and at the far detectors. The
detector considered in this first example is a modest one, i.e., a
15~kton liquid argon TPC and we assume 6 years of neutrino and 2 years
of antineutrino data taking in the near detector. All in all, the
total number of years of neutrino and antineutrino running would be
14, 6 with only the far detector plus 8 with both detectors
(\textbf{ScI}). The degeneracies related to the wrong choice of the
atmospheric mass squared difference are again depicted with cyan
contours. Notice that the second detector by itself does not give any
better results than the far detector. Synergy effects when combining
the data of both detectors have a remarkable outcome. This is shown
in the bottom panel, where this combination is depicted with black
contours. It is clear from the results shown in Fig.~\ref{fig:0.09}
that the combination of these two setups (\nova as the far detector
plus a 15 kton liquid argon as the near detector with the same $L/E$
at both sites) could resolve the neutrino mass hierarchy if $\sin^2 2
\theta_{13}>0.09$. Let us notice that in Fig.~\ref{fig:0.09} we have
only illustrated the case for $\delta=285^{o}$ because this is the
value of $\delta$ which gives the worst resolution of
$sgn(\deltaunotre)$ for the value of $\sin^2 2\theta_{13}$
assumed~\footnote{The value of $\delta$ assumed in the rest of the
  figures is the one which gives the worst resolution of the type of
  neutrino mass hierarchy for each value of $\sin^2
  2\theta_{13}$.}. Hence, it is important to realize that, when one  
explores the full range of $\delta$, there exist some values of the
CP--violating phase $\delta$ for which the sign of the atmospheric
mass difference could be resolved also in a more modest scenario, i.e,
a scenario with fewer years of data taking, or a scenario with a
smaller detector, or a combination of both cases. It might certainly
happen that Nature is kind enough to have chosen a different value for
the CP--violating phase. In that case, the determination of the type
of neutrino mass hierarchy would be much easier. Here we present the
most pessimistic case. On the other hand, another experiment which may
improve the results depicted in Fig.\ref{fig:0.09} is MINOS, which
could be sensitive to $\sin^2 2 \theta_{13} > 0.08$ if the total
number of protons on target achieves a value of $16 \times
10^{20}$~\cite{MINOS}.

\begin{figure}[t]
\begin{center}
\epsfig{file=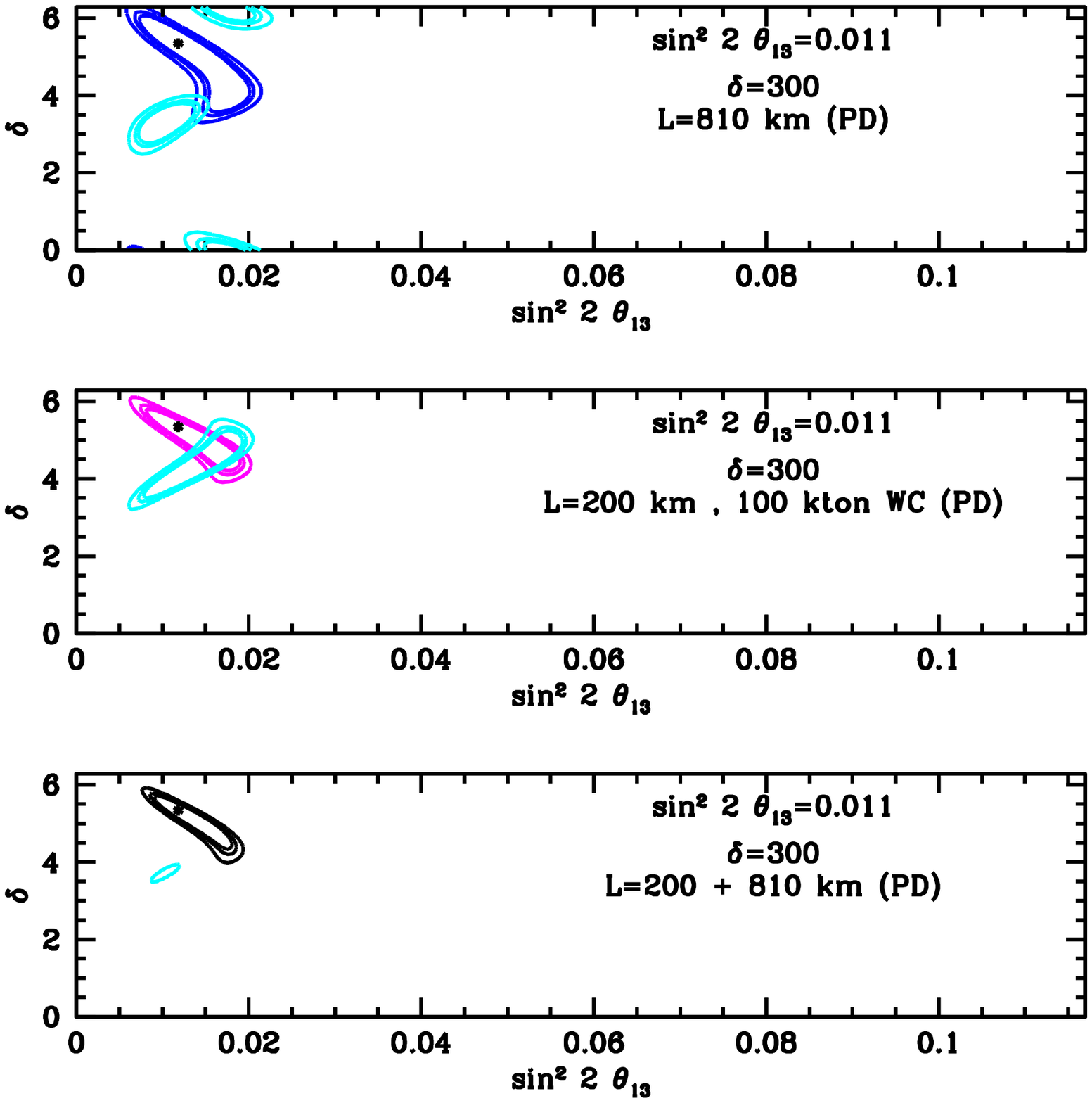,width=12.0cm}
\caption{\textit{The same as in Fig.~(\ref{fig:0.019pd}) but for
    $\sin^2 2 \theta_{13}=0.011$ and $\delta=300^{\circ}$.}}
\label{fig:0.01pd}
\end{center}
\end{figure}

The next step in the study we are presenting here is to assume that
the true value of $\sin^2 2 \theta_{13}$ is smaller than the one
considered in Fig.~\ref{fig:0.09}, but still within the sensitivity
reach of the \novasp experiment. We show in Fig.~\ref{fig:0.07} the
results for the same combination of two detectors with the same $L/E$
(that is, \nova far detector and a 15~kton liquid argon TPC near
detector) for $\sin^2 2 \theta_{13}=0.076$. The plots are
qualitatively very similar to those of Fig.~\ref{fig:0.09}, but as is
evident from the bottom pannel, in this case, when combining the data
at the two detectors, the fake solution remains present, although only
at the $99\%$ C.L. In this case, the value of $\delta$ for which the
resolution of the type of neutrino mass hierarchy is most difficult is
$\delta = 300^{o}$. Again, let us point out that this happens for this
specific value of $\delta$. Hence, for any other value of the
CP--violating phase, the results would be even better.

If $\sin^2 2 \theta_{13}<0.07$, in order to eliminate the fake
solutions, it is necessary to increase the mass of the near
detector. We show the results for a true value of $\sin^2 2
\theta_{13}=0.058$ in Fig.~\ref{fig:0.058} and $\delta = 300^{o}$, 
where we have upgraded the modest 15 kton near detector to a 50 kton
water-\v{C}erenkov detector ($M_{\rm efc} = 13.5$ and 35~kton,
respectively) and for $\sin^2 2 \theta_{13}= 0.043$ and $\delta =
315^{o}$ in Fig.~\ref{fig:0.043}, where the near detector has been
upgraded to a 100~kton water-\v{C}erenkov detector ($M_{\rm efc} =
70$~kton). We should notice here that for other values of the
CP--violating phase $\delta$ a smaller detector would provide the
statistics required for resolving the type of hierarchy. This will be
illustrated below by the use of exclusion plots in the ($\sin^2 2
\theta_{13}$, $\delta$) plane. 

\begin{figure}[t]
\begin{center}
\begin{tabular}{ll}
\hskip -0.5cm
\epsfig{file=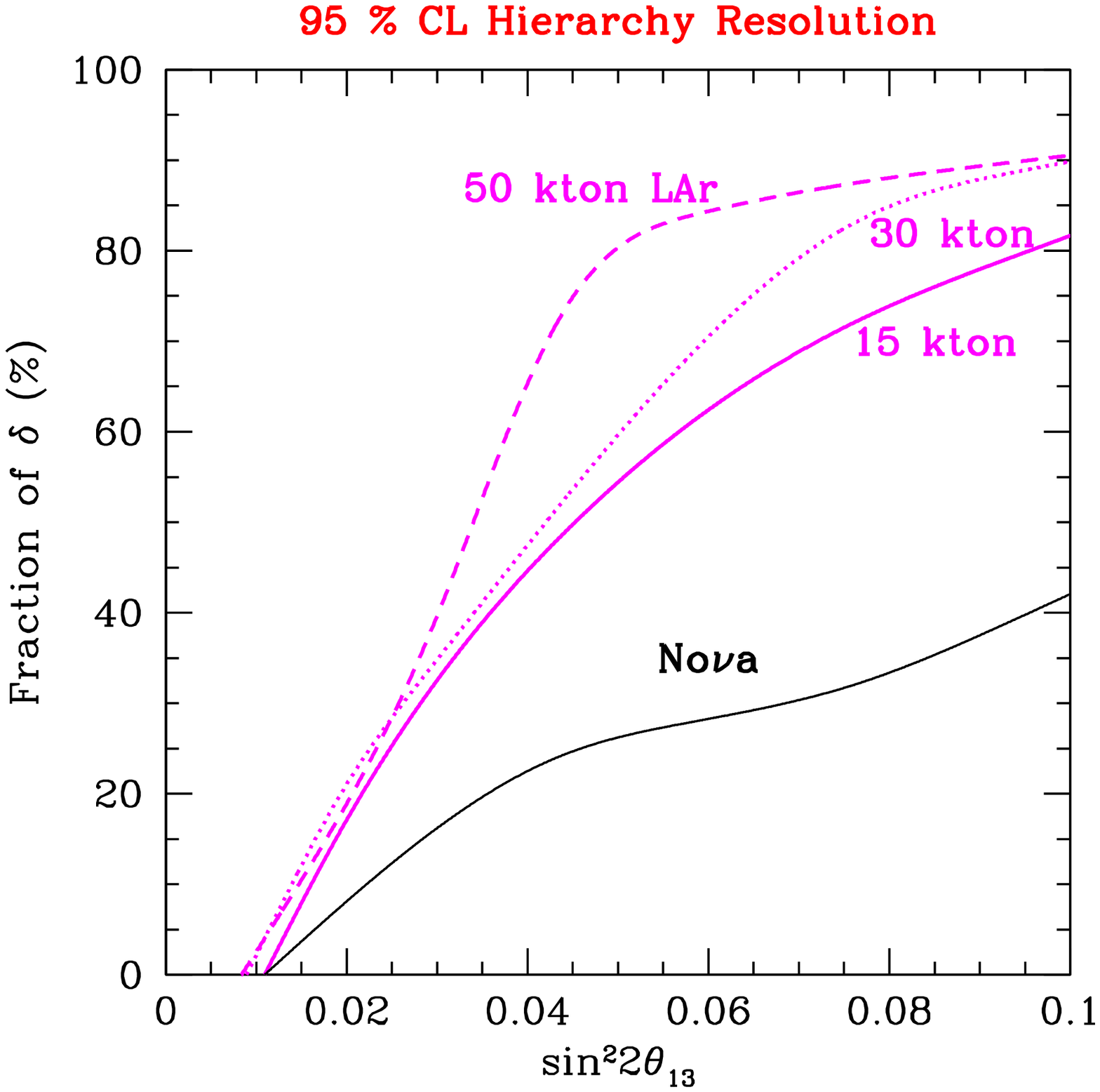, width=8.1cm} &
\hskip 0.cm
\epsfig{file=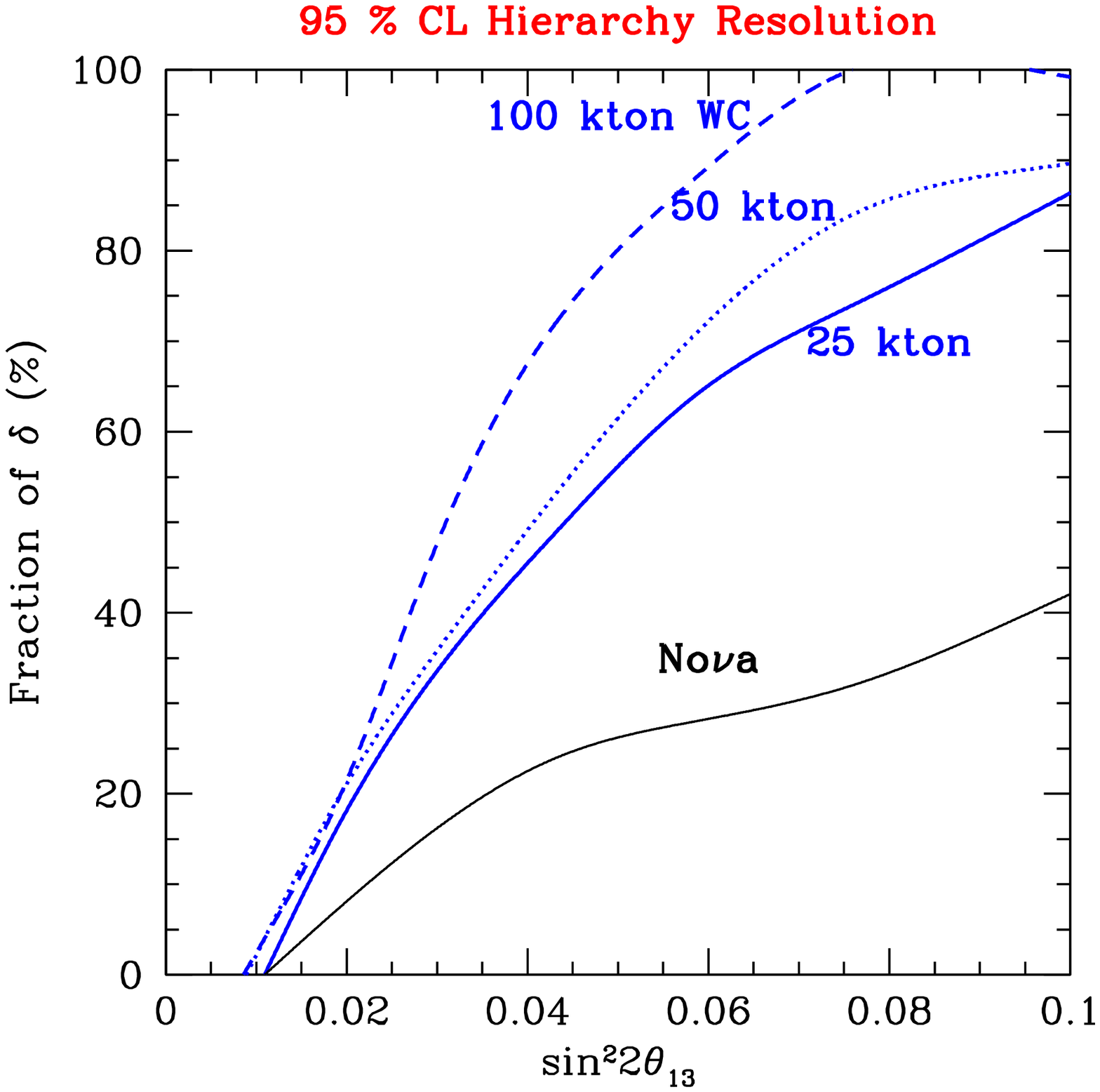, width=8.1cm} \\
\hskip 3.2truecm
{\small (a)}            &
\hskip 3.8truecm
{\small (b)}
\end{tabular}
\end{center}
\caption{\textit{The fraction of values of $\delta$ as a function
    of $\sin^2 2\theta_{13}$ for which the resolution of the neutrino
    mass hierarchy at 95\% C.L. is achieved, assuming that Nature
    has chosen $\deltaunotre$ to be positive but the analysis of
    the data is performed with the opposite sign. The number of
    protons on target per year is $6.5\times 10^{20}$ and the
    statistics considered here consists of 9 years of neutrino running
    plus 5 years of antineutrino running for the far detector and 6
    years of neutrino running plus 2 years of antineutrino running for
    the near site (\textbf{ScI}). The total number of years of running
    would be 14.The lower solid black line corresponds to the case of
    the proposed \nova far detector alone. (a) For liquid argon
    detectors in the near site: 15~kton (upper solid magenta line),
    30~kton (dotted magenta line) and 50~kton (dashed magenta
    line). (b) For water-\v{C}erenkov detectors at the near site:
    25~kton (upper solid blue line), 50~kton (dotted blue line) and
    100~kton (dashed blue line).}}
\label{fig:excl1}
\end{figure}

As we have summarized in the description of the experimental
configuration, \textbf{ScI} does not include a proton driver. It turns 
out that for vales of $\sin^2 2 \theta_{13}< 0.04$ the efficient mass
of the detector required would exceed $M_{\rm efc} = 70$~kton which
might be a too demanding solution. A more practical approach to
resolve the $sgn(\deltaunotre)$-degeneracies would be to upgrade the
experimental setup from \textbf{ScI} to scenario II (\textbf{ScII}, 
with a Proton Driver), which might constitute a more feasible option
that would provide the statistics required for small values of $\sin^2
2 \theta_{13}$. We present the results for $\sin^2 2 \theta_{13} =
0.019$ and $\delta = 300^{o}$ in Fig.~\ref{fig:0.019pd}. The top panel
shows the results of the data at the far detector, which within the
\textbf{ScII} scheme would take data for 6 years in the neutrino plus
4 years in the antineutrino channel. The medium panel shows the
results for the near detector, a 100~kton water \v{C}erenkov detector,
which would take data for 3 years in the neutrino channel plus 1 year
in the antineutrino channel. The total number of years of running in
the \textbf{ScII} strategy would be 10. We can see from the bottom
panel that for this scenario, $sgn(\deltaunotre)$ could be determined
at $99\%$ C.L. for $\sin^2 2\theta_{13} \simeq 0.02$ for any value of
the CP--violating phase $\delta$. Finally, we show in
Fig.~\ref{fig:0.019pd}, the same combination of near and far detectors
as in Fig.~\ref{fig:0.019pd} and with a proton driver, but for $\sin^2
2 \theta_{13} = 0.011$ and and $\delta = 300^{o}$. It gives similar
results, but as can be noticed from the bottom pannel, when adding the
data at both detectors, the fake solution remains at the 99 $\%$ C.L.
 
\begin{figure}[t]
\begin{center}
\begin{tabular}{ll}
\hskip -0.5cm
\epsfig{file=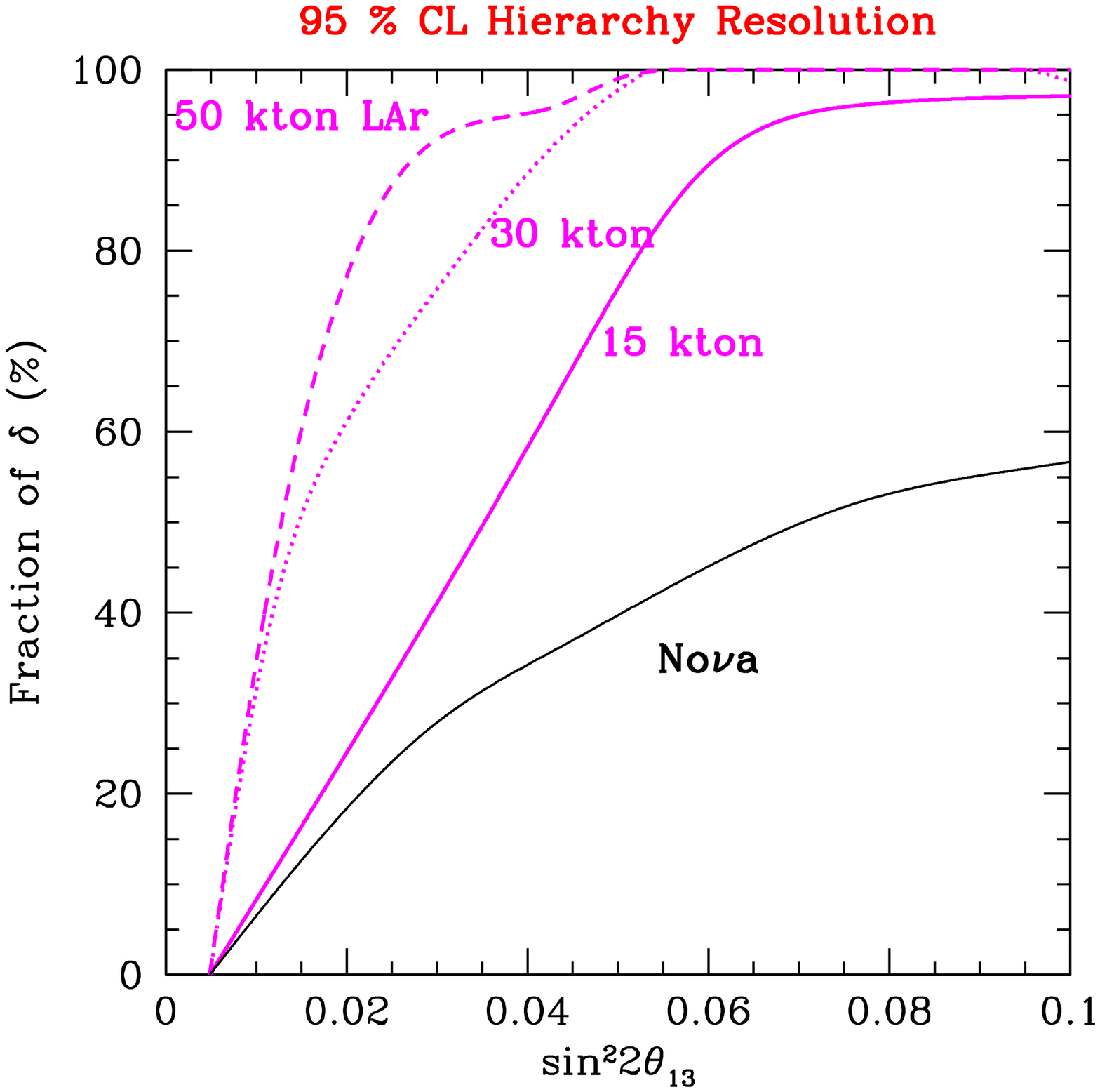, width=8.1cm} &
\hskip 0.cm
\epsfig{file=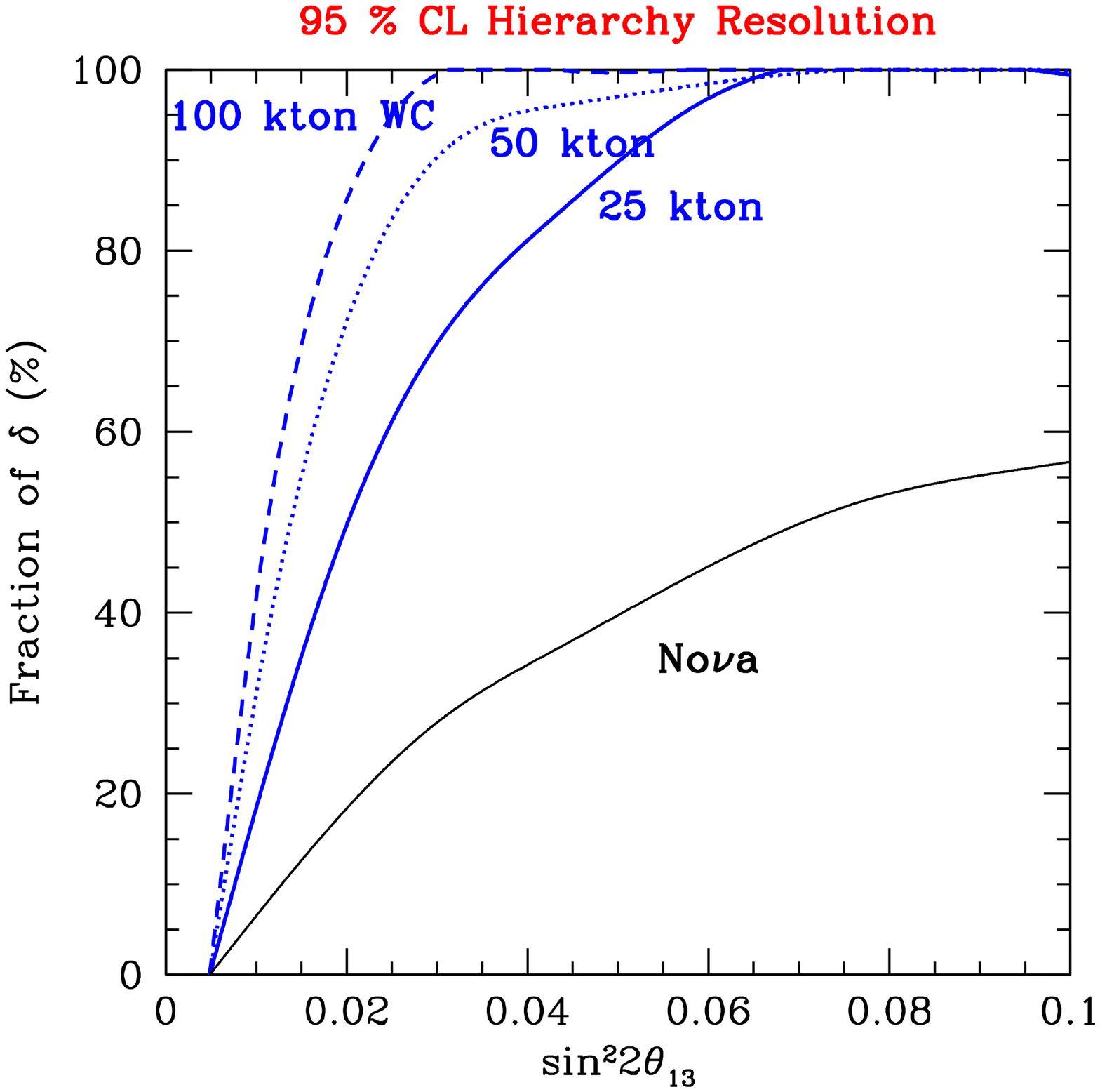, width=8.1cm} \\
\hskip 3.2truecm
{\small (a)}            &
\hskip 3.8truecm
{\small (b)}
\end{tabular}
\end{center}
\caption{\textit{The same as Fig.~(\ref{fig:excl1}) but for
    scenario II (\textbf{ScII}), with a proton driver. The number of
    protons on target per year in the proton driver scenario is
    $25 \times 10^{20}$. We have considered 6 years of neutrino
    running and 4 years of antineutrino running in the proposed \nova
    far detector and 3 years of neutrino running and 1 year of
    antineutrino running in the near detector. The total number of
    years of running would be 10.
}}
\label{fig:excl2}
\end{figure}

We summarize these results with the exclusion plots in
Figs.~\ref{fig:excl1} and \ref{fig:excl2}, where we show the fraction
of $\delta$ as a function of $\sin^2 2\theta_{13}$, for which the sign
of the atmospheric mass squared difference can be determined at 95\%
C.L. in \textbf{ScI} and \textbf{ScII}, respectively. The left panel
of both figures shows the results for liquid argon  detectors for
three different masses: 15~kton (upper solid magenta line), 30~kton
(dotted magenta line) and 50~kton (dashed magenta line). The right
panel shows the results for the case of three water-\v{C}erenkov
detectors: 25~kton (upper solid blue line), 50~kton (dotted blue line)
and 100~kton (dashed blue line). In each panel it is also shown what
would be the result if there was no second detector and all data was
collected by the proposed \nova far detector (lower black curve). It
is evident from the plots that adding the second detector helps
considerably  and that even a 15~kton liquid argon detector would
improve the capabilities of the experiment enormously. It is also
clear from the figures that if $\sin^2 2\theta_{13} < 0.04-0.05$ the
use of a proton driver is crucial. Running with a proton driver and
with a second detector, placed at the same $L/E$ as the far detector,
would make the determination of the type of mass hierarchy feasible
for values of $\sin^2 2\theta_{13}$ as small as 0.02 and for a
substantial range of values of the CP--violating phase $\delta$. 

\begin{figure}[t]
\begin{center}
\begin{tabular}{ll}
\hskip -0.5cm
\epsfig{file=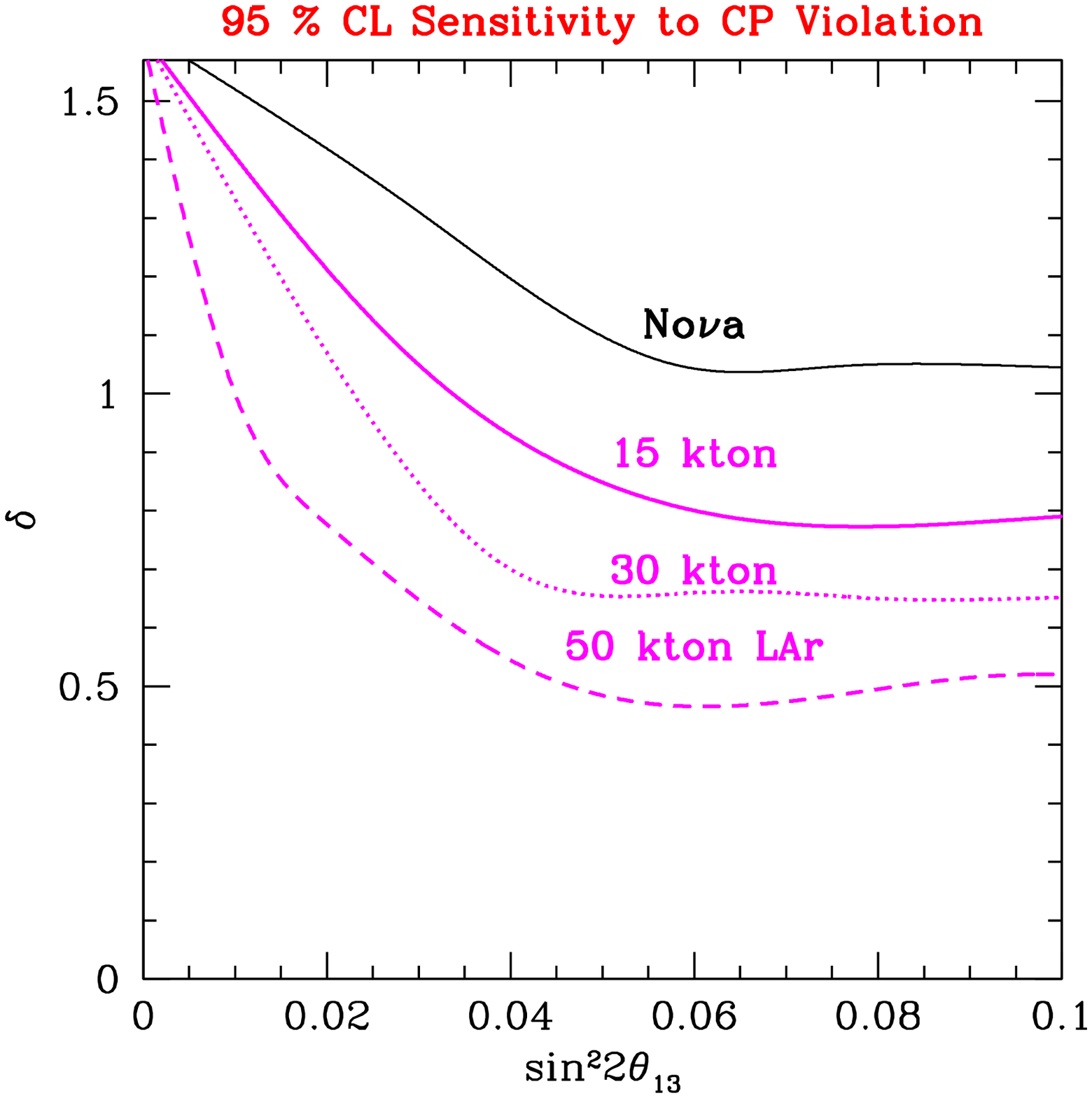, width=8.1cm} &
\hskip 0.cm
\epsfig{file=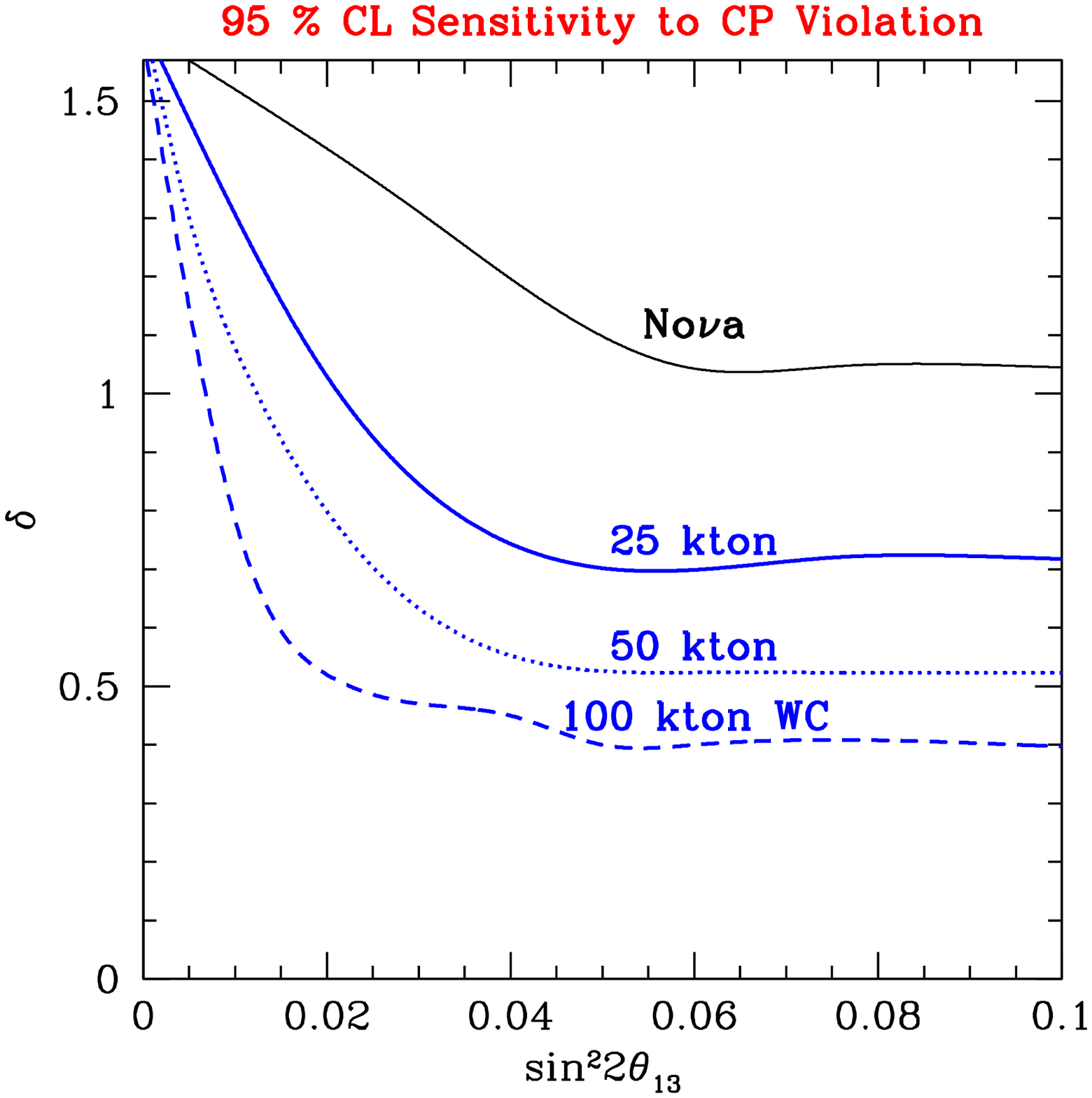, width=8.1cm} \\
\hskip 3.2truecm
{\small (a)}            &
\hskip 3.8truecm
{\small (b)}
\end{tabular}
\end{center}
\caption{\textit{Sensitivity to the determination of CP--violation at
    95\% C.L. as a function of $\sin^2 2\theta_{13}$ for the first
    quadrant of $\delta$. It is shown the minimum value of $\delta$ at
    which the error at $95\%$ C.L. reaches $\delta=0$, and therefore
    the CP--violating case is indistinguishable from the CP--conserving
    one, i.e, $\delta=0$. (a) For liquid argon detectors: 15~kton
    (lower solid magenta line) , 30~kton (dotted magenta line) and
    50~kton (dashed magenta line). (b) For water-\v{C}erenkov
    detectors at the near site: 25~kton (lower solid blue line),
    50~kton (dotted blue line) and 100~kton (dashed blue line). Also
    depicted is the result if there was no second detector and all
    data was collected by the proposed at \nova far detector (upper
    solid black line).}}
\label{fig:exclCP}
\end{figure}

Once we have studied in detail the extraction of the type of the
neutrino mass spectrum in our approach, we analyze the reach in
searching for CP--violation. In this case, we have only considered
\textbf{ScII}, as a larger neutrino flux is required. As can be seen
by comparing the top and middle panels with the bottom panel in
Figs.~\ref{fig:0.09}-~\ref{fig:0.01pd}, resolving the mass hierarchy
removes one of the existing degeneracies, increasing the sensitivity
to CP--violation. Whereas without the second detector almost none of
the values of $\delta$ can be excluded (making the determination of
CP--violation challenging) by adding the second detector, only the
region related to the true hierarchy is left. To be quantitative, we
show in Fig.~\ref{fig:exclCP} as a function of $\sin^2 2\theta_{13}$,
the minimum value of $\delta$ at which the error at $95\%$
C.L. reaches $\delta=0$, and therefore the CP--violating case is
indistinguishable from the CP--conserving one, i.e, $\delta=0$. Below
the curves it is impossible to tell the difference between the
CP--violating and the CP--conserving cases, whereas above them
CP--violation could be stated at 95\% C.L. As in Figs.~\ref{fig:excl1}
and \ref{fig:excl2}, the left panel depicts the results for three
liquid argon detectors and in the right panel, those for the case of
three water-\v{C}erenkov detectors are plotted. In each panel it is
also shown what would be the result if there was no second detector
and all data was collected by the proposed \nova far detector (upper
black curve). As an indication, we have only shown the results for the
first quadrant of $\delta$~\footnote{The exclusion lines are symmetric
  in $\delta$ in   the vacuum case.}. From the figure we see that for
values of $\sin^2 2\theta_{13} < 0.04$ the sensitivity to
CP--violation decreases quite fast, which is related to the fact that
the mass hierarchy cannot be resolved for a larger range of values of
$\delta$. Nevertheless, it is evident from the figure that if
$\theta_{13}$ is large enough, by using the proposed configuration,
CP--violation could be established, e.g, for the modest case of a
15~kton liquid argon second detector if the true $\delta \gtap 45^{o}$
and $\sin^2 2\theta_{13} \gtap 0.04$.

Finally, let us comment that throughout this work we are assuming
maximal mixing in the atmospheric sector. If this turns out not to be
the true solution, another degeneracy would show up due to the
inability to distinguish the octant where the mixing angle
$\theta_{23}$ lies. This is due to the fact that current atmospheric
and long-baseline neutrino experiments are sensitive to $\sin^2
2\theta_{23}$, but not to $\sin^2 \theta_{23}$. This degeneracy can be
broken by future atmospheric neutrino experiments~\cite{futatm} or by
neutrino factories making use of both golden and silver
channels~\cite{CDGGCHMR00,silver}. We expect this degeneracy not to
change sustantially our conclusions with respect to the determination
of the type of neutrino mass hierarchy, but we do expect a reduction
in the sensitivity to CP-violation. In any case, the study of this
degeneracy is beyond the scope of this work.

\subsection{Dependence on  $|\deltaunotre|$}

\begin{figure}[t]
\begin{center}
\epsfig{file=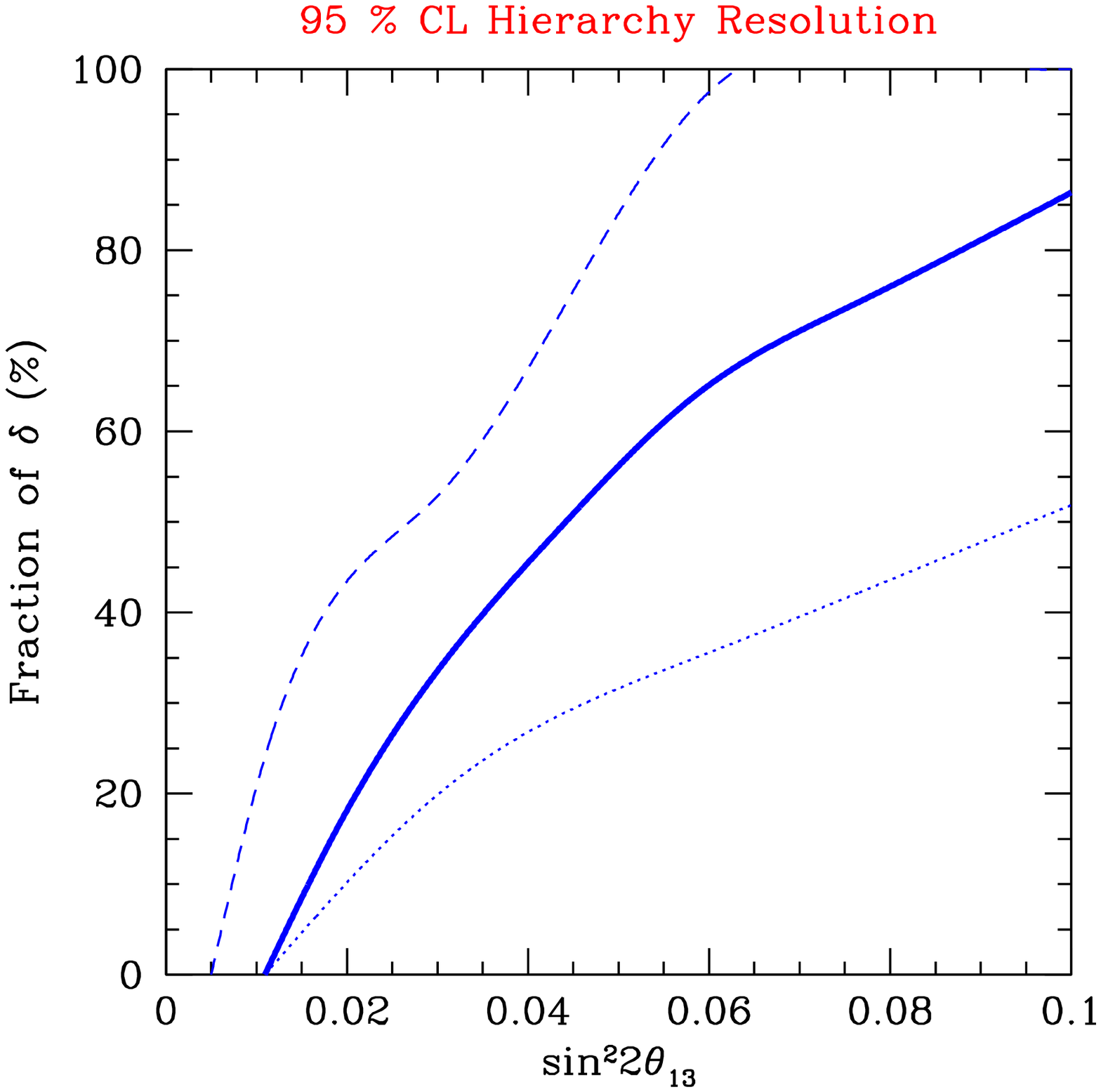,width=9.0cm} 
\caption{\textit{Sensitivity to $sgn(\deltaunotre)$ for different
    values of $|\deltaunotre|$ = $2.0 \times 10^{-3} \ \rm{eV}^2$
    (dotted line), $2.4 \times 10^{-3} \ \rm{eV}^2$ (solid line), and
    $3.0 \times 10^{-3} \ \rm{eV}^2$ (dashed line). We have exploited
    the data from two off-axis detectors located at 200~km and 810~km
    by assuming \textbf{ScI} with a 25~kton water-\v{C}erenkov
    detector at the near site and the proposed \nova detector at the
    far site.}}
\label{fig:difdm}
\end{center}
\end{figure}

Throughout this work we have been assuming a fixed value for
$|\deltaunotre| = 2.40 \times 10^{-3} \ \rm{eV}^2$. However, the
value of this parameter is known currently with a precision of $\sim
30\%$ at 90\% C.L.~\cite{K2K,SKatm}. In addition, from
Eq.~(\ref{eq:probdifffull}), we can see that the value of the
asymmetry increases monotonically as the vaccum oscillation phase
increases, and therefore the asymmetry is larger for larger
$|\deltaunotre|$. This correspondingly means better sensitivity to the
type of mass hierarchy the larger the atmospheric mass square
difference is. Hence, although a precision at the level of $\sim 5\%$
is expected to be achieved by the time this experiment could turn
on~\cite{newNOvA,messierp04}, it is very important to investigate the
effect of a different value for $|\deltaunotre|$ on the results
presented above.

We have performed such a study, in a similar way to Ref.~\cite{MPP05}.
In Fig.~\ref{fig:difdm} the sensitivity to the sign of
$\deltaunotre$ for different values of $|\deltaunotre|= (2.0
$; $2.4$; $3.0) \times 10^{-3} \rm{eV}^2$. We have assumed
\textbf{ScI} with a 25~kton water-\v{C}erenkov detector at 200~km. We
depict the fraction of $\delta$ for which the type of neutrino mass
hierarchy can be determined as a function of $\sin^2 2\theta_{13}$,
for three different values of $|\deltaunotre|$ = $2.0 \times 10^{-3}
\rm{eV}^2$ (dotted line); $2.4 \times 10^{-3} \rm{eV}^2$ (solid line)
and $3.0 \times 10^{-3} \rm{eV}^2$ (dashed line). As anticipated, the
larger the value of $|\deltaunotre|$, the better the sensitivity to
$sgn(\deltaunotre)$. As for comparison, for a value of
$|\deltaunotre|$ = $2.0 \times 10^{-3} \rm{eV}^2$, the results are
only slightly better than what is achieved with just the far \nova
detector if $|\deltaunotre|$ = $2.4 \times 10^{-3}
\rm{eV}^2$. Conversely, if $|\deltaunotre|$ = $3.0 \times 10^{-3}
\rm{eV}^2$, the sensitivity is at the level of that for
$|\deltaunotre|$ = $2.4 \times 10^{-3} \rm{eV}^2$ with four times more
statistics, i.e., a 100~kton water-\v{C}erenkov detector instead. On
the other hand, let us point out that since the CHOOZ~\cite{CHOOZ}
bound is weaker for small values of $|\deltaunotre|$, the loss in
range for $\theta_{13}$ is not as large as one would na\"{\i}vely
think from Fig.~\ref{fig:difdm}.

Hence, if future atmospheric and long-baseline experiments determine
that the actual value of $|\deltaunotre|$ happens to be smaller than
the one assumed for this work, a different solution must be considered
in order to achieve a comparable sensitivity. A possible solution
would be to adopt a larger $L/E$, which could be accomplished either by
considering longer baselines or larger off-axis distances, i.e.,
smaller energies. Another possibility would be to consider running
with the low-energy configuration of the NuMI beam. Nevertheless, these
modified experimental setups would imply a reduction of the neutrino
flux at the detectors, which would require a detailed analysis to
evaluate their actual capabilities. On top of this, if $\theta_{13}$
is very small and $|\deltaunotre|$ is also small, then the
construction of the proton driver and possibly a longer neutrino
running would be necessary.

\section{Comparing different locations for the second detector}
\label{diffloc}

\begin{figure}[t]
\begin{center}
\epsfig{file=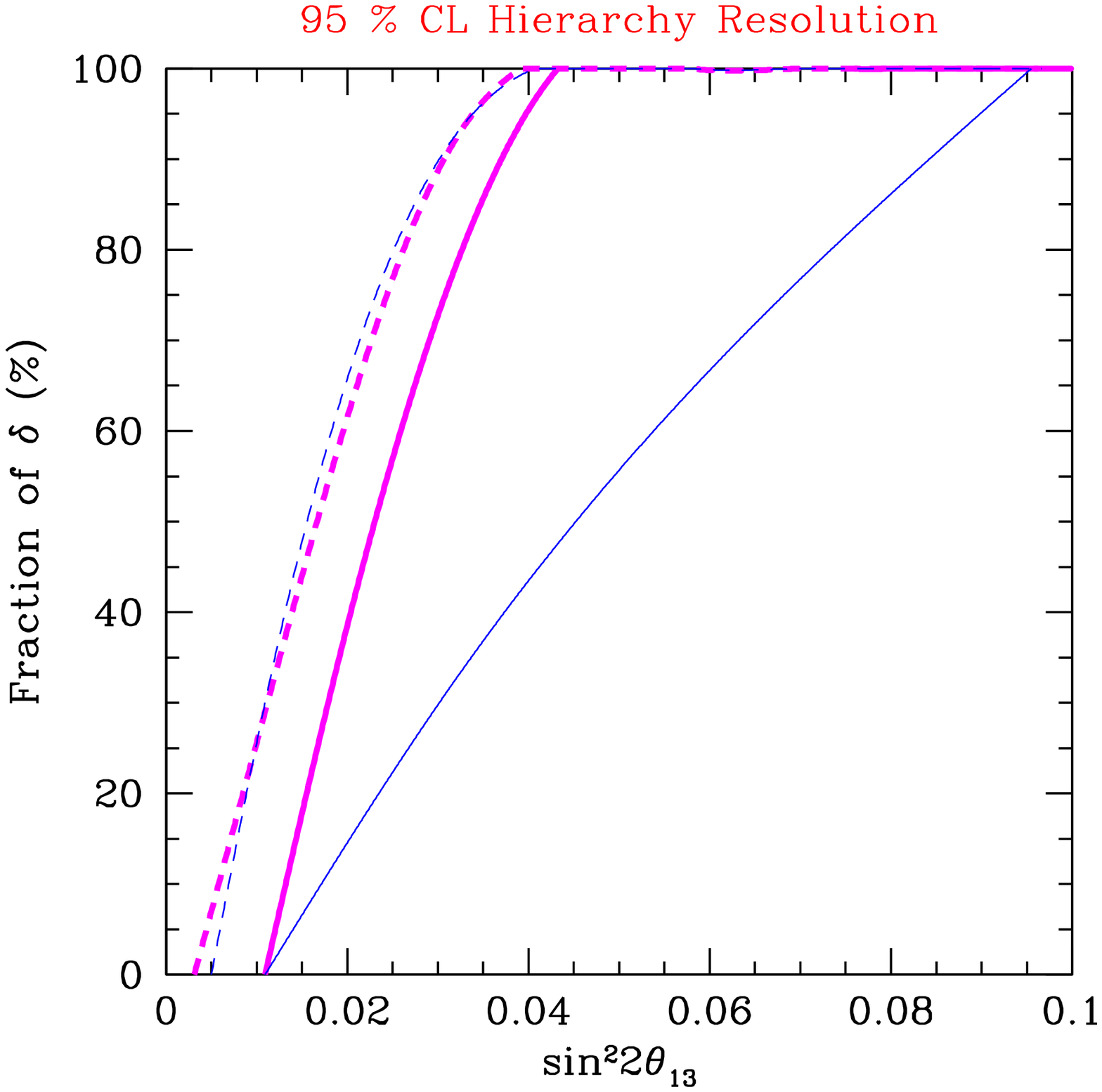,width=9.0cm}
\caption{\textit{Comparison of the capabilities for determining
    $sgn(\deltaunotre)$ for two different setups for the second
    detector: at the same $L/E$ as the far detector (thick lines) and
    at the second oscillation maximum (thin lines). For illustrative
    purposes, we have assumed two 50~kton liquid argon TPC detectors, 
    both at the far and at the near sites (see Ref.~\cite{MPP05})
    without proton driver. We have considered 5 years of neutrino
    running (solid lines) and 5 years of neutrino plus 5 years of
    antineutrino running (dashed lines).}}
\label{fig:second}
\end{center}
\end{figure}

In the previous section we have studied the capabilities for
determining the type of neutrino mass hierarchy and of CP violation by
adding a second detector to the already proposed far detector for the
\nova configuration. The results obtained are for a Super-\novasp-like
configuration~\cite{MPP05}, that is with the second detector placed
such that (the peak energy) has the same $L/E$ of the far
detector. This choice is based on the findings of
Refs.~\cite{HLW02,MNP03} and the study of Ref.~\cite{MPP05}, where it
was shown that such a configuration is specially sensitive to matter
effects even with only the neutrino run. 

Interestingly, in the \nova proposal~\cite{newNOvA}, the possibility
to add a second detector is pointed out. The approach there consists
in placing the second detector at the second oscillation maximum,
where the matter effect is smaller by a factor of three because the
energy is smaller by the same factor. However, for such a
configuration there is no cancellation between the vacuum terms at
both detectors, and the CP--violating effects are also different
(larger by a factor of three in the second detector for the same
reason as above). These facts imply that the matter-dependent terms
are of the same order of those carrying information on CP--violation,
which makes much more difficult to disentangle the type of mass
hierarchy from CP--violating effects. Therefore, a measurement using
only a neutrino run would not be as sensitive to the type of neutrino
mass spectrum as in the case of Super-\nova and a running also in the
antineutrino mode would be needed in order to achieve a comparable
sensitivity. In addition, going to the second oscillation maximum
means going more off-axis, and hence losing flux. In this case the
loss in number of events amounts to about a factor of 15 with respect
to the first detector~\footnote{Compare with the Super-\nova case
  where in the second detector the number of events is just about half
  that at the far detector~\cite{MPP05}.}.

In Fig.~\ref{fig:second}, we compare these two possibilities for
placing the second detector. For illustrative purposes, we have
assumed for this comparison the same type of detectors as for
Super-\novasp~\cite{MPP05}, i.e., two 50 kton liquid argon TPC. The
results depict the cases without proton driver, where solid lines are
for 5 years of neutrino running and dashed lines are for 5 years of
neutrino plus 5 years of antineutrino running. As was anticipated
above, the configuration for which both detectors are located such
that the vacuum oscillation phase is the same (thick solid magenta
lines), i.e., the same $L/E$, is much better when running only with
neutrinos than that with the second detector at the second oscillation
maximum (thin solid blue lines). In order to achieve comparable
sensitivities for the determination of the type of neutrino mass
hierarchy, the antineutrino run is necessary for the case of \nova
plus a second detector located at the second oscillation maximum (thin
dashed blue line). On the other hand, in the case of Super-\nova the
antineutrino information does not add any synergy, but just statistics
(thick dashed magenta line). Hence, it is important to note that in
order to determine $sgn(\deltaunotre)$ with the second detector at the
same $L/E$, the antineutrino run is not really needed. 

As it is well known, provided $\sin^2 2\theta_{13}$ is in the range
accesible to conventional neutrino beams, the unique contribution of
the NuMI neutrino program, and in particular of the \nova
experiment~\cite{newNOvA}, will be the resolution of the type of 
neutrino mass hierarchy. Thus, from the results presented in
Fig.~\ref{fig:second}, we learn that the best approach in case the
second detector is needed would be the one studied in
Ref.~\cite{MPP05} and throughout the present paper.

\section{Conclusions}
\label{conclusions}

Determining the type of neutrino mass hierarchy, whether normal or
inverted, constitutes one of the fundamental questions in neutrino
physics. Future long-baseline experiments aim at addressing this
fundamental issue but suffer typically from degeneracies with other
CP--conserving and CP--violating parameters, namely $\theta_{13}$,
$\delta$ and $\theta_{23}$. The presence of such degeneracies limits
the sensitivity to the type of hierarchy as well as to
CP--violation. Many studies focus on resolving such degeneracies,
e.g. by combining more than one
experiment~\cite{BMW02,HLW02,MNP03,otherexp,mp2,MP05}, using more
than one detector~\cite{BNL,MN97,BCGGCHM01,silver,BMW02off,MPP05}, or
using information from atmospheric neutrino data~\cite{HMS05}. 

In the present article, we follow the strategy delined in
Ref.~\cite{MPP05}, in which the type of hierarchy is determined free
of degeneracies in only one neutrino experiment with two off-axis
detectors and by using the neutrino beam alone. The off-axis
configurations are chosen in such a way that at the two detectors the
ratio $L/E$ is the same, and so the vacuum oscillation
phases. Comparing the probabilities of neutrino conversion at the two
distances, the vacuum term (hence the dominant CP--violating one)
cancels out in the difference~\cite{MNP03,MPP05}. The normalized
difference $\probdiff$ depends only on matter effects and its sign is
determined by the type of neutrino mass hierarchy. CP--violating terms
can give relevant contributions only for small values of $\sin^2
2\theta_{13}$. This implies that the determination of the type of
neutrino mass spectrum, exploiting this method, suffers no
degeneracies from other parameters. In the Super-\nova proposed
configuration,  two off-axis 50~kton liquid argon detectors, one at a
far site and one at a shorter distance $\sim 200$~km, were used. Even
if the construction of small liquid argon detectors has proven to be
possible, it might be difficult to build the required large liquid
argon detectors on the \nova timescale. Here we have considered a more
realistic proposal, in which at the far site the \nova currently
proposed detector~\cite{newNOvA} is used. The second detector should
be added off-axis at about 200~km. We study both liquid argon and
water-\v{C}erenkov tecniques and different sizes for the detectors.
Even if establishing the type of hierarchy with this setup does not
require antineutrinos, here we consider both neutrino and antineutrino
running modes and we analyze the capabilities of determining
CP--violation as well. 

We have proposed to sequence the experiment. In the first phase, the
experiment would have just the far off-axis detector as in the \nova
proposal~\cite{newNOvA}. In the second phase, the near off-axis
detector is added at 200~km. We have considered two different
scenarios. In scenario I (\textbf{ScI}) the experiment runs with the 
NuMI beam for 6 years as in the NO$\nu$A proposal and then with a
second off-axis detector 6 years in the neutrino mode and 2 with
antineutrinos. In scenario II (\textbf{ScII}) the beam is upgraded
by the use of a proton driver and the number of years of data taking
with both detectors is reduced to 3 and 1 for neutrinos and
antineutrinos, respectively. We have performed an $\chi^{2}$ analysis
of the simulated data on the ($\sin^{2} 2 \theta_{13}$, $\delta$)
plane. In Figs.~\ref{fig:0.09}--\ref{fig:0.01pd}, for different values
of $\sin^2 2\theta_{13}$ and different type and mass of detectors, we
show the capabilities of different experimental setups to measure
$\sin^2 2 \theta_{13}$, $\delta$ and determine the type of
hierarchy. The values of $\delta$ chosen are those that give the worst 
results, so they represent the most pessimistic case. In particular,
for each set of parameters, we depict the sensitivity when each
detector, far and near off-axis, is considered alone, and we show how
the degeneracies can be lifted when the data at the two locations are
combined. Obviously, the smaller the value of $\sin^2 2 \theta_{13}$,
the larger the required effective mass of the second detector. In
Figs.~\ref{fig:excl1} and \ref{fig:excl2}, we compare the reach of 
different choices for the second detector, for \textbf{ScI} and
\textbf{ScII}, respectively. We depict the fraction of values of
$\delta$ as a function of $\sin^2 2 \theta_{13}$ for which the type of
hierarchy can be determined at 95\%~C.L. For \textbf{ScI}, we see that
adding a modest detector, such as a 15~kton liquid argon or 25~kton
water-\v{C}erenkov detector, would yield a substantial
improvement with respect to the performance with just the proposed
\nova detector. For example, the fraction of $\delta$ values is  $\sim
85\%$ ($\sim 65\%$) at $\sin^2 2 \theta_{13} = 0.1 (0.06)$, while in
the case of the \nova configuration alone such fraction decreases to
$\sim 40\%$ ($\sim 25\%$). The construction of a larger detector, such
as a 100~kton water-\v{C}erencov one, would have even a more dramatic
impact, raising the fraction of $\delta$ to 100\% for $\sin^2 2
\theta_{13} \gtap 0.075$ and to $70\%$ even for $\sin^2 2 \theta_{13}
\gtap 0.04$. In \textbf{ScII}, the higher flux would allow the
determination of $sgn(\deltaunotre)$, {\em independently of
  $\delta$}, for $\sin^2 2 \theta_{13} \gtap 0.06$ even for a small
detector. More remarkable results would be obtained for larger
detectors or longer data-taking time. 

Thanks to the years of antineutrino run, the discussed experimental
setup would also allow to search for CP-violation. We show the
sensitivity to CP-violation at 95\%~C.L. in Fig.~\ref{fig:exclCP} for
different types of second off-axis detector. Obviously, also in the
case of CP--violation, the sensitivity increases with statistics and
improves greatly with respect to the proposed \nova experiment.

From our analysis it is clear that the choice of building the second
detector should be guided by the knowledge of the value of $\sin^2 2
\theta_{13}$, provided by \nova itself and by other
experiments~\cite{MINOS,futurereactors,T2K}. If $\sin^2 2 \theta_{13}<
0.01$--$0.02$, no sensitivity on the type of mass spectrum could be
reached and the construction of a second detector would not improve it
significantly. On the contrary, if $\sin^2 2 \theta_{13}\gtap 0.02$, a
sufficiently large second detector would {\em guarantee} the
determination of the type of hierarchy (almost) independently of the
value of $\delta$. The actual type and mass of the second off-axis
detector should be chosen on the basis of the measured value of
$\sin^2 \theta_{13}$. As the sensitivity to $sgn(\deltaunotre)$
depends also on the unknown value of $\delta$, the best strategy would
be to build a detector for which the determination of the type of
hierarchy can be achieved for a large fraction of values of the
CP-violating phase, e.g. 60\%--70\%. In case we are lucky and Nature
has chosen a favorable value of $\delta$, the above discussed number
of years of data taking, if not less, would allow to answer this
question. Otherwise additional years of neutrino run would be needed
to resolve this issue. 

In addition, as discussed in Ref.~\cite{MPP05}, the reach of this
experimental setup depends on the value of $|\deltaunotre|$. In
Fig.~\ref{fig:difdm} we  also show the sensitivity to
$sgn(\deltaunotre)$ for three values of $|\deltaunotre| = (2.0
$; $2.4$; $3.0) \times 10^{-3} \rm{eV}^2$. The sensitivity decreases
for lower values of $|\deltaunotre|$ and larger statistics might be
required if $|\deltaunotre|$ lies in the low side of the presently
allowed range~\cite{SKatm,suzukitaup}, depending on the true value of
$\delta$. 

On the other hand, in the recent \nova proposal the possibility of
adding a second off-axis detector at the second maximum has been
considered. Let us notice that the flux at the second location is
greatly suppressed by the large off-axis angle, though. In our
proposal the reduction in flux is not as dramatic as the second
detector should be located at a much shorter baseline, $\sim
200$~km. In Fig.\ref{fig:second}, we compare the capabilities of the
two experimental setups in determining the type of
hierarchy. Considering 5 years of only neutrino running, the
Super-\novasp-like configuration discussed in the present article
could resolve the neutrino mass ordering for any value of $\delta$ for
$\sin^2 2 \theta_{13} \gtap 0.04$. The other experimental setup, with
the second detector at second oscillation maximum, cannot fully lift
the degeneracies for $\sin^2 2 \theta_{13} < 0.095$. The reach of the
two setups become comparable only when additional 5 years of
antineutrino running are considered. Having in mind that the unique
contribution of the \nova experiment is to determine the type of
neutrino mass spectrum, we consider that the the best approach in case
the second detector is needed would be the one studied in
Ref.~\cite{MPP05} and throughout the present paper. 

In conclusion, following the strategy illustrated in ref.~\cite{MPP05},
we have studied different experimental setups for a sequenced off-axis
experiment which could achieve considerably better sensitivity to the
type of neutrino mass hierarchy with respect to the proposed \nova
experiment. The problem of degeneracies is weakened and for a
sufficiently large second off-axis detector completely solved. The
capabilities of measuring the value of the CP--violating phase
$\delta$ have also been analyzed in detailed and have been shown to be
greatly improved. Let us stress again that, for $\sin^2 2 \theta_{13}$
in the range accesible to conventional neutrino beams, the main
breakthrough of the NuMI neutrino program, and in particular of the
\nova experiment~\cite{newNOvA}, will be the resolution of the type of
neutrino mass hierarchy. This goal should be achieved in the most
efficient and fast possible way. Here we have presented a very
powerful approach to this remarkable problem.

\section{Acknowledgments}
We would like to thank A. Para, S. Parke and A. Rubbia for
discussions. SPR thanks the Theory Division at CERN and the Department
of Theoretical Physics of the University of Valencia for hospitality
during his stay. SP is also grateful for the hospitality of the
Fermilab Theoretical Physics Department during her visit. Our
calculations made extensive use of the Fermilab General-Purpose
Computing Farms~\cite{farms}. SPR is supported by NASA Grant
ATP02-0000-0151 and by the Spanish Grant FPA2002-00612 of the
MCT. Fermilab is operated by URA under DOE contract DE-AC02-76CH03000.

\end{document}